\documentclass[final,times,twocolumn]{elsarticle}

\usepackage{xcolor}
\usepackage{soul}

\usepackage[english]{babel}
\usepackage{amsmath}
\usepackage{graphicx}
\usepackage{dcolumn}
\usepackage{bm}

\newcommand{\bXi}{\boldsymbol{\Xi}}

\newcommand{\bSigma}{\boldsymbol{\Sigma}}
\newcommand{\bx}{\mathbf{x}}
\newcommand{\by}{\mathbf{y}}
\newcommand{\bM}{\mathbf{M}}
\newcommand{\bq}{\mathbf{q}}
\newcommand{\bC}{\mathbf{C}}
\newcommand{\bK}{\mathbf{K}}
\newcommand{\bJ}{\mathbf{J}}

\newcommand{\be}{\begin{equation}}
\newcommand{\ee}{\end{equation}}
\newcommand{\bea}{\begin{eqnarray}}
\newcommand{\eea}{\end{eqnarray}}

\newcommand{\Tr}{{\rm Tr}}

\usepackage{hyperref}
\usepackage{pdfpages}

\journal{Chaos, Solitons \& Fractals}

\begin{document}

\begin{frontmatter}

\title{From Chaos to Synchrony in Recurrent Excitatory--Inhibitory Networks with Target-Specific Inhibition}

\author[inst1]{Carles Martorell\corref{cor1}}
\ead{carlesma@onsager.ugr.es}
\cortext[cor1]{Corresponding author}

\author[inst1]{Rub\'en Calvo}
\author[inst2]{Alessia Annibale}
\author[inst1]{Miguel A. Mu\~noz}

\address[inst1]{Departamento de Electromagnetismo y F\'isica de la Materia and Instituto Carlos I de F\'isica Te\'orica y Computacional, Universidad de Granada, E-18071 Granada, Spain}

\address[inst2]{Department of Mathematics, King's College London, SE11 6NJ London, United Kingdom}

\begin{abstract}
Biological neural networks can operate in qualitatively distinct dynamical regimes, and transitions between these regimes are thought to underlie changes in computation and behavior. A seminal theoretical foundation for understanding such regimes was provided by Sompolinsky, Crisanti, and Sommers (SCS), who showed that random recurrent networks undergo a transition from quiescent activity to asynchronous chaos. This result established a paradigmatic link between random connectivity, dynamical instability, and internally generated fluctuations in neural circuits. Dynamic mean-field theory (DMFT), introduced in this context, has since become a powerful framework for constructing phase diagrams of recurrent network dynamics.
Here, we extend the SCS framework to two-population firing-rate networks with segregated excitatory and inhibitory neurons and target-specific inhibitory couplings that break excitation–inhibition balance. Using DMFT, we derive self-consistent equations for macroscopic observables and stability conditions that distinguish mean-driven from fluctuation-driven routes to instability. We identify three qualitatively distinct organizations of phase diagrams. In inhibition-dominated and strictly balanced networks, the system displays the classical SCS phenomenology: quiescent activity and asynchronous chaos. In excitation-dominated networks, additional regimes emerge, including persistent activity. Depending on the eigenvalues of the stability matrix, excitation-dominated networks can further exhibit either chaos with non-vanishing mean activity, corresponding to structured or synchronous chaos, or collective oscillations. However,  these oscillations do not coexist with chaotic fluctuations around the periodic mean trajectory; instead, the onset of collective oscillations suppresses the chaotic component, in a regime reminiscent of input-induced suppression of chaos.
Together, these results generalize the seminal SCS theory to circuits with explicit excitatory–inhibitory structure and target-specific inhibition, providing a unified DMFT-based framework linking inhibitory architecture to the organization of large-scale dynamical regimes in recurrent neural networks.
\end{abstract}

\end{frontmatter}

\section{Introduction}

Understanding how macroscopic dynamical regimes emerge from disordered recurrent networks is a central problem at the interface of statistical physics and theoretical neuroscience. Neural circuits display a broad repertoire of dynamical regimes, ranging from irregular asynchronous activity to coherent oscillations and low-dimensional structured dynamics \cite{Kandel_principles_2000,dayan_theoretical_2005, gerstner_neuronal_2014,izhikevich_dynamical_2007,stringer_high-dimensional_2019}. These regimes arise primarily from the interplay between the connectivity of excitatory (E) and inhibitory (I) populations and the intrinsic nonlinear dynamics of neurons and synapses, and are thought to support distinct computational functions.

From the perspective of theoretical neuroscience, a central challenge is to determine how minimal structural ingredients ---namely, structural disorder and, conversely, low-rank (structured) connectivity--- organize the dynamical phase diagram and delimit transitions between dynamical regimes.

A major step in this direction was provided by Sompolinsky, Crisanti, and Sommers (SCS) \cite{sompolinsky_chaos_1988}, who showed that large random recurrent networks undergo a transition from a stable fixed point to a regime of high-dimensional chaotic activity as synaptic heterogeneity
increases.  This transition can be characterized analytically in the thermodynamic limit using dynamical mean-field theory (DMFT), providing a paradigmatic example of a disorder-induced phase transition into a high-dimensional chaotic state. In the literature, this fluctuation-driven state is commonly referred to as asynchronous chaos 
\cite{harish_asynchronous_2015,dahmen_second_2019,li_tuning_2020, martorell_ergodicity_2025}.
Subsequent work extended this framework to more realistic settings, incorporating diverse nonlinear transfer functions, heterogeneous connectivity, and additional biological constraints, thereby revealing a broader diversity of dynamical regimes
\cite{martorell_ergodicity_2025, kadmon_transition_2015,schuecker_optimal_2018,crisanti_path_2018,landau_coherent_2018,dick_linking_2024,martorell_dynamically_2024,kadmon_efficient_2025}.

Within this framework, we focus on the role of connectivity structure, which can be naturally described as low-rank components superimposed on random interactions. 
In particular, a non-zero mean connectivity provides a simple example of a rank-one addition to purely random networks, inducing collective states with non-vanishing mean activity ---analogous to symmetry-broken phases in simple models of magnetism \cite{sherrington_solvable_1975, nishimori_statistical_2001}--- that may coexist with chaos \cite{martorell_ergodicity_2025, schuecker_optimal_2018,crisanti_path_2018,martorell_dynamically_2024,landau_coherent_2018,  fournier_high-dimensional_2025, fournier_non-reciprocal_2025}.

A crucial step toward biological plausibility is the explicit segregation of neurons into excitatory (E) and inhibitory (I) populations, which naturally introduces a rank-two structure in the mean connectivity. In this setting, the rank-two component encodes the mean interactions between the two populations. When inhibition is sufficiently strong, excitatory and inhibitory inputs can dynamically balance each other, giving rise to asynchronous chaotic activity \cite{van_vreeswijk_chaos_1996,vreeswijk_chaotic_1998,brunel_dynamics_2000,renart_asynchronous_2010,buendia_jensens_2019,mastrogiuseppe_intrinsically-generated_2017}. Away from this balanced regime, however, the specific E/I structure can generate collective modes that shape the macroscopic dynamics of the system \cite{mastrogiuseppe_linking_2018,schuessler_dynamics_2020,beiran_shaping_2021,dubreuil_role_2022}.

This naturally raises the following question: can rank-two networks ---such as those appearing naturally in E/I networks--- sustain symmetry-broken chaotic states, and more specifically --in analogy with the rank-one perturbations--- can they exhibit an oscillatory mean activity coexisting with chaos? Within the E/I framework, this issue remains only partially understood. Most previous studies have focused either on balanced regimes and the emergence of asynchronous chaotic activity \cite{harish_asynchronous_2015, garcia_del_molino_synchronization_2013,mastrogiuseppe_intrinsically-generated_2017}, or on more general E/I structures without specifically addressing the dynamical interplay between oscillatory collective modes and chaotic dynamics \cite{mastrogiuseppe_linking_2018,kadmon_transition_2015}.

It is a well-established biological fact  that inhibitory neurons do not act uniformly across the network, but instead target excitatory and inhibitory populations in distinct ways, giving rise to structured asymmetries in synaptic strength and connectivity patterns \cite{kepecs_interneuron_2014,tremblay_gabaergic_2016,corral_lopez_excitatory-inhibitory_2022}.
Such target-specific inhibition is thought to play an important role in shaping network activity and regulating stability and gain \cite{dayan_theoretical_2005}. 

Motivated by the previous observation, in this work, we study an extension of the SCS framework with explicit E/I populations, each endowed with a non-zero mean interaction and with target-specific inhibitory couplings. The resulting network combines Gaussian disorder with a minimal rank-two structured component, allowing us to analyze how collective modes induced by the E/I architecture interact with the disordered background. Using a two-population dynamical mean-field theory (DMFT), we derive self-consistent equations for the mean activities and autocorrelations, together with stability criteria that distinguish disorder-driven from mean-driven instabilities.

Our results show that target-specific inhibition acts as a key regulator of the phase diagram. In particular, the relative strength of the two inhibitory couplings defines a phase boundary that separates two qualitatively distinct regimes within the intermediate region between the asynchronous-chaotic and fixed-point phases. On one side of this boundary, the system may develop a chaotic state with non-vanishing mean (broken symmetry); on the other, it exhibits coherent oscillations.  However,  chaotic dynamics cannot be sustained on top of an oscillatory collective mean: the onset of coherent oscillations suppresses high-dimensional chaotic fluctuations. Accordingly, the transition from asynchronous chaos to coherent oscillations reflects a competition between a disorder-driven chaotic state and a low-dimensional collective mode, with the latter dominating beyond the phase boundary. Overall, our main contribution is to demonstrate that target-specific inhibition determines which collective instability --chaotic or oscillatory-- becomes dominant, thereby qualitatively reorganizing the phase diagram.

The remainder of the paper is organized as follows. Sec.~2 introduces the two-population model and its parametrization; Sec.~3 presents the DMFT equations; Sec.~4 characterizes the non-quiescent phases and the corresponding phase diagrams; Sec.~5 analyzes chaotic and oscillatory dynamics using Lyapunov and Kuramoto measures; and Sec.~6 discusses the implications of our results and possible extensions.

\section{Model definition}

We consider a firing-rate network with two populations, excitatory (E) and inhibitory (I), each of size $N$. The firing rates at time $t$ of excitatory neuron $i$ and inhibitory neuron $j$ are denoted by $x_i(t)$ and $y_j(t)$, respectively, with $i,j = 1,\dots,N$. 
Defining the activity vectors as \( \mathbf{x}(t) = [x_1(t), \dots, x_N(t)]^T \) and \( \mathbf{y}(t) = [y_1(t), \dots, y_N(t)]^T \), their dynamics follow a set of coupled differential equations \cite{dayan_network_2005, vogels_neural_2005},
\begin{equation}
 \partial_t 
 \left[\begin{array}{c}
       \bx(t)  \\
       \by(t)
  \end{array}\right]  = -\left[\begin{array}{c}
       \bx(t)  \\
       \by(t)
  \end{array}\right] + \phi\left( g \, \mathcal{W}\left[\begin{array}{c}
       \bx(t)  \\
       \by(t)
  \end{array}\right] \right) 
    \label{eq:dynamics-def}
\end{equation}
where $\partial_t$ denotes the temporal derivative, 
$\phi(x)$ is a gain function that we set to $\phi(x) = \tanh(x)$ and $g$ is a coupling-strength parameter. 
\footnote{Three remarks are in order. First, although the model studied here differs from the original SCS model, the two can be mapped onto each other, as shown explicitly in \cite{martorell_dynamically_2024}. Second, we use the term ``activity'' for the state variable; since it can take negative values, it should be understood not as a firing rate, but as an abstract fluctuation around a baseline activity level \cite{, dayan_theoretical_2005, gerstner_neuronal_2014}. Third, the case of a general model with unequal population sizes is discussed in the Supplemental Material \cite{martorell_supplemental_nodate}.}

The synaptic connectivity between elements in the two populations is encoded in the matrix 
$\mathcal{W}$,
\begin{equation} \label{Eq. W}
    \mathcal{W}= \left[\begin{array}{cc}
       W^{EE}  & \; -\beta W^{EI} \\
       W^{IE}  & -\delta W^{II}
    \end{array} \right]
\end{equation}
whose entries \(W^{AB}_{ij}\) represent the synaptic weight from neuron \(j\) in population \(B\) to neuron \(i\) in population \(A\). Unless stated otherwise, each submatrix \(W^{AB}\) is drawn independently from a fully asymmetric Gaussian ensemble with i.i.d.  entries of mean \(J_0/N\) and variance \(J^2/N\). \footnote{This approach replaces biologically plausible sparse connections, which preserves Dale's principle, by fully connected random matrices with independent entries. The validity of this substitution is justified in the thermodynamic limit ($N \to \infty)$, where both connectivity schemes become statistically equivalent, as demonstrated in the Supplementary Material \cite{martorell_supplemental_nodate}.} The positive parameters \(\beta,\delta>0\) control the strength of inhibitory couplings, providing a minimal extension of the SCS model that independently tunes E/I balance through target-specific inhibition.

In DMFT one focuses on the instantaneous mean activities and two-time autocorrelation functions. These macroscopic observables are defined as:
\begin{equation}
\begin{aligned}
    \hat{M}_x(t) &=& \left \langle \frac{1}{N} \sum_{i=1}^N  x_i(t) \right \rangle_\mathcal{W},   \\
    \hat{M}_y(t) &=& \left \langle \frac{1}{N} \sum_{i=1}^N  y_i(t) \right \rangle_\mathcal{W},  
    \label{Eq. hat M}
\end{aligned}
\end{equation}
and 
\begin{equation}
\begin{aligned}
\hat{C}_x(t, s) &= \left \langle \frac{1}{N} \sum_{i=1}^N  x_i(t)\,  x_i(s) \right \rangle_\mathcal{W},   \\
\hat{C}_y(t, s) &= \left \langle \frac{1}{N} \sum_{i=1}^N  y_i(t) \,  y_i(s)\right \rangle_\mathcal{W},  
\label{Eq. hat C}    
\end{aligned}
\end{equation}
where  \( \langle \cdot \rangle_\mathcal{W} \) denotes averaging over network realizations (that is, ensemble average).
We do not need to consider cross-correlations ---defined by $\langle x_i(t) y_j(t) \rangle_\mathcal{W}$---  since different neurons turn out to become fully uncorrelated in the thermodynamic limit ($N \longrightarrow \infty$), as noted below in Section \ref{Subsec. DMF}. 
\footnote{Observe that for some parameter values the dynamics become oscillatory, breaking strict stationarity (time-translation invariance). As a result, macroscopic observables such as those in Eqs. (\ref{Eq. hat M}) and (\ref{Eq. hat C}) retain explicit time dependence. Unlike the stationary one-population SCS model (where self-averaging holds), some DMFT standard techniques cannot be carried directly on the E/I case.}

The mean inputs to excitatory and inhibitory neurons are respectively:
\begin{equation} \label{Eq. mean input}
    \begin{aligned}
\hat{\mu}_x(t) &=& 
\hat{M}_x(t) - \beta \hat{M}_y(t),
\\ 
\hat{\mu}_y(t) &=& 
\hat{M}_x(t) - \delta \hat{M}_y(t).
    \end{aligned}
\end{equation} 
A system is said to be in a \emph{balanced state} when excitatory and inhibitory inputs cancel on average at all times, so that $\hat{\mu}_x(t) = \hat{\mu}_y(t) = 0$, and the dynamics are entirely fluctuation-driven \cite{kadmon_transition_2015, van_vreeswijk_chaos_1996, brunel_dynamics_2000}. 
As we will see, this naturally arises in inhibition-dominated regimes, $1 < \delta <\beta$; and, in particular, under the \textit{strict balance condition}, \( \beta = \delta = 1 \).

\section{Dynamical Mean-Field Description} \label{Subsec. DMF}
As profusely established in the literature, systems of the form in Eq.~(\ref{eq:dynamics-def}) effectively decouple in the $N \longrightarrow \infty$ limit, so that each neuron is described by a  single-unit stochastic process driven by a deterministic mean field plus auto-correlated residual network fluctuations \cite{sompolinsky_chaos_1988,kadmon_transition_2015,crisanti_path_2018,schuecker_optimal_2018,martorell_dynamically_2024,martorell_ergodicity_2025,metz_dynamical_2025}. 

As shown in detail in Supplemental Material using the Martin–Siggia–Rose generating-functional formalism, in the specific case of segregated excitatory and inhibitory populations, the DMFT approach maps the full network dynamics onto an effective two-dimensional stochastic system, with one equation per population \cite{martorell_supplemental_nodate}: 
\begin{equation}
\begin{aligned}
\partial_t x(t) &=& -x(t) + \phi\; \big(gJ_0\,\mu_x(t) + gJ\,\eta_x(t)\big),\\
\partial_t y(t) &=& -y(t) + \phi\; \big(gJ_0\,\mu_y(t) + gJ\,\eta_y(t)\big),
\label{Eq: DMF-beta}
\end{aligned}
\end{equation}
where the effective mean inputs, which are the DMFT counterparts of Eq. (\ref{Eq. mean input}), are expressed as 
\begin{equation}
\begin{bmatrix}
    \mu_x(t) \\
    \mu_y(t)
\end{bmatrix} 
= \mathcal{N}_{\beta, \delta} 
\begin{bmatrix}
    M_x(t) \\
    M_y(t)
\end{bmatrix}, 
\label{Eq. mus} 
\end{equation}
where $M_{x/y}(t)$ are the mean values of $x$ and $y$ respectively and the  effective rank-two coupling matrix is
\begin{equation}
\mathcal{N}_{\beta, \delta} = 
\begin{bmatrix}
    1 & - \beta \\
    1 & - \delta
\end{bmatrix}.
\label{Eq. N} 
\end{equation}
The stochastic fields \(\eta_x(t)\), \(\eta_y(t)\) are independent, zero-mean Gaussian noises with covariance functions
\begin{equation}
\begin{bmatrix}
    \Delta_x(t, s) \\
    \Delta_y(t, s)
\end{bmatrix} = \mathcal{M}_{\beta, \delta} \begin{bmatrix}
    C_x(t, s) \\
    C_y(t, s)
\end{bmatrix}, 
\label{Eq. deltas} 
\end{equation}
with
\begin{equation}
\mathcal{M}_{\beta, \delta} = 
\begin{bmatrix}
    1 &  \quad \beta^2 \\
    1 &  \quad \delta^2
\end{bmatrix}.
\label{Eq. M} 
\end{equation}
The $C_{x,y}(t,s)$ functions are two-time autocorrelation functions, the DMFT counterpart of Eq. (\ref{Eq. hat C}). These equations need to be closed self-consistently through the macroscopic order parameters, 
\begin{eqnarray}
 M_x(t) = \langle x(t)\rangle, &  M_y(t)  = \langle y(t)\rangle \label{Eq. M def} \\  
 C_{x}(t,s) =\langle x(t)\,x(s)\rangle, \; & C_y (t,s) = \langle y(t)\,y(s)\rangle \label{Eq. C def}
\end{eqnarray}
where \(\langle\cdot\rangle\) denotes averages over noise realizations $\eta_{x,y}$.
\footnote{We note that the effective neurons $x(t)$ and $y(t)$ obey Eq.~(\ref{Eq: DMF-beta}), an uncoupled stochastic system driven by independent noises $\eta_x(t)$ and $\eta_y(t)$. Although $\mu_x,\mu_y$ and $\Delta_x,\Delta_y$ are defined self-consistently from the statistics of $x(t)$ and $y(t)$, the stochastic contributions come from independent processes, so the cross-correlation for effective neurons factorizes (see \cite{martorell_supplemental_nodate} for details):
\begin{equation}
\langle x(t) y(s) \rangle = M_x(t)M_y(s).
\end{equation}
Thus, in the infinite-size limit, excitatory and inhibitory neurons are uncorrelated.}

Let us first note that in the symmetric case, $\beta = \delta$, Eqs.~(\ref{Eq: DMF-beta}) describe the dynamics of two statistically identical, but independent subsystems that differ only in their noise realizations (see \ref{App. DMF balance condition}). As a result, both populations can be described separately by the same one-dimensional DMFT equation, for their corresponding activity variable ($x(t)$ or $y(t)$) that can be called, generically $z(t)$:
\begin{equation} \label{Eq: DMF SCS}
\partial_t  z(t) = -z(t)+  \phi \big( g J_0 \mu_z(t) + g J \, \eta(t) \big),
\end{equation}
such that $\langle \eta(t) \eta(s) \rangle = \Delta_z(t, s)$, with 
\begin{eqnarray}
\mu_z(t) &=& (1 - \beta)\, M_z(t), \\
\Delta_z(t, s) &=& (1 + \beta^2) \, C_z(t, s).
\end{eqnarray}
where $M_z(t) = \langle z(t) \rangle$ and $C_z(t, s) = \langle z(t) z(s) \rangle$, averaged over $\eta(t)$ realizations. 
Thus, Eq.~(\ref{Eq: DMF SCS}) is identical to the DMFT equation of the one-population model with non-zero mean coupling ---i.e. the extension of the SCS model \cite{sompolinsky_chaos_1988} analyzed in \cite{martorell_dynamically_2024,martorell_ergodicity_2025}--- upon the parameter mapping 
\begin{equation}
J^{\rm{eff}}_0 =  (1 - \beta) J_0 , \quad
J^{\rm{eff}} = \sqrt{1 + \beta^2}\, J.
\end{equation}
In the strictly balanced case, \(\beta=\delta=1\), the effective mean coupling vanishes, 
$J^{\rm{eff}}_0 = 0$. The mean input therefore vanishes,
\(\mu_z(t)=0\), yielding a balanced, purely fluctuation-driven state
equivalent to the zero-mean SCS model \cite{sompolinsky_chaos_1988,
martorell_dynamically_2024, martorell_ergodicity_2025}.

\section{Phase Diagram and Phase Transitions of the Excitatory–Inhibitory Model} \label{Sec. Dynamical regimes}

Building on the DMFT equations, we analyse the phase diagram in terms of the coupling parameters $J_0$, $J$, and $g$, and determine how the phase structure depends on the target-specific inhibitory couplings $(\beta,\delta)$. We first study fixed-point solutions and their stability, and then extend the analysis to oscillatory and chaotic regimes.

\subsection{Quiescent-state linear-stability analysis}

Within the DMFT framework, the mean activities are self-consistently determined by Eqs. (\ref{Eq. M def}), yielding the corresponding fixed-point solutions:
\begin{equation}
\begin{aligned}
M_x &= \int_{-\infty}^{\infty} \mathcal{D}u \, \phi\Big(gJ_0(M_x - \beta M_y) + g J\sqrt{q_x + \beta^2 q_y}u  \Big), \\
M_y &= \int_{-\infty}^{\infty} \mathcal{D}u\,  \phi\Big(gJ_0(M_x - \delta M_y) + g J\sqrt{q_x + \delta^2 q_y}u \Big), 
\end{aligned}\label{Eq. M_beta}
\end{equation}
where $\mathcal{D}u = du / \sqrt{2\pi} e^{-\frac{u^2}{2}}$ represents the Gaussian measure and $q_x$ and $q_y$ denote the corresponding fixed-point autocorrelations. From Eq. (\ref{Eq. C def}), the fixed-point autocorrelations are self-consistently determined by
\begin{equation}
\begin{aligned}
q_x &= \int_{-\infty}^{\infty} \mathcal{D}u\,  \phi\Big(gJ_0(M_x - \beta M_y) + g J\sqrt{q_x + \beta^2 q_y}u \Big)^2,  \\
q_y &= \int_{-\infty}^{\infty} \mathcal{D}u \,  \phi\Big(gJ_0(M_x - \delta M_y) + g J\sqrt{q_x + \delta^2 q_y}u \Big)^2.
\end{aligned}  
\label{Eq. q_beta}
\end{equation}

\begin{figure*}[tbh]
    \centering

    \begin{minipage}[t]{0.40\textwidth}
        \vspace{0pt}
        \centering
        \includegraphics[width=0.90\linewidth]{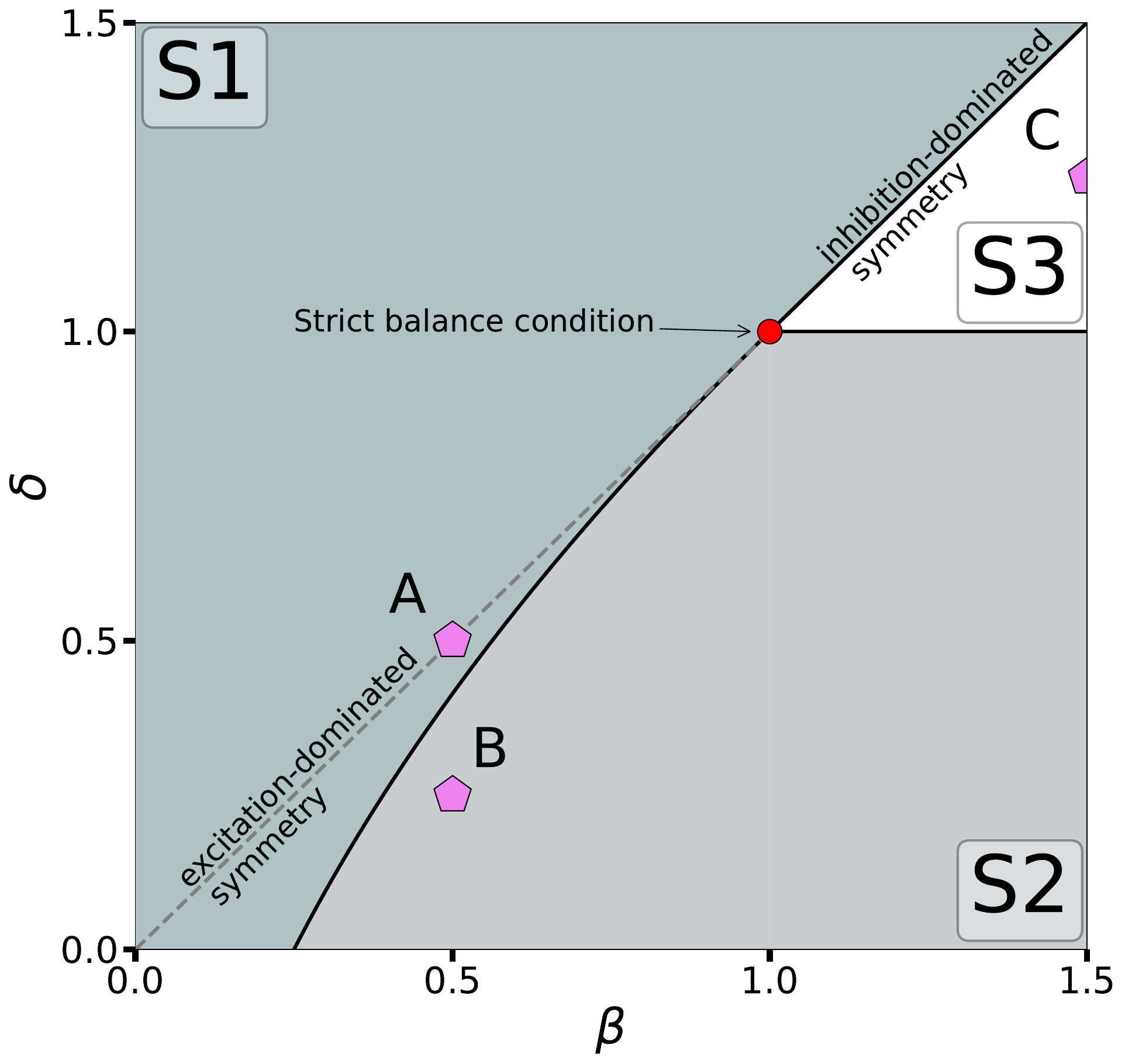}
    \end{minipage}
    \hfill
    \begin{minipage}[t]{0.55\textwidth}
        \vspace{20pt}
        \centering
        \small
        \renewcommand{\arraystretch}{1.15}
        \begin{tabular}{p{0.035\linewidth} p{0.40\linewidth} p{0.40\linewidth}}
            \multicolumn{3}{c}{\textbf{Leading eigenvalue of $\mathcal{N}_{\beta,\delta}$}} \\
            \hline \\[-1.8ex]
            &\textit{Condition} & \textit{Nature} \\
            \hline \\[-1.8ex]

            \textbf{\normalsize S1. } & $\delta < 1$ and $(1+\delta)^2 - 4\beta > 0$; or $\delta > 1$ and $\delta > \beta$ & Real positive eigenvalue
            \\[0.8ex]

            \textbf{\normalsize S2. } &$\delta < 1$ and $(1+\delta)^2 - 4\beta < 0$ &  Complex eigenvalue with positive real part \\[0.8ex]

            \textbf{\normalsize S3. }  &$1 < \delta < \beta$ & Complex eigenvalue with negative real part 
             \\

            \hline
        \end{tabular}
    \end{minipage}

\caption{
\textbf{Nature of the leading eigenvalue of $\mathcal{N}_{\beta,\delta}$ ---determining the type of resulting phase diagram--- across the $(\beta,\delta)$ plane.} The table on the right summarizes the classification of the leading eigenvalue into three distinct cases, denoted by S1, S2, and S3. 
The corresponding conditions associated to these cases split the $(\beta,\delta)$ parameter space into the three regions as shown in the left panel. 
In this panel, solid lines denote the analytical boundaries between these regimes, while the dashed line identifies the symmetric-inhibition condition, $\beta=\delta$. 
Violet pentagons identify three representative parameter choices analyzed in Fig.~\ref{fig: phase_space}.
}
    \label{fig:1}
\end{figure*}

The DMFT equations admit a trivial quiescent fixed point given by \(M_x = M_y = 0\) and \(q_x = q_y = 0\), whose stability can be assessed by linearizing Eqs. (\ref{Eq. M_beta}) and (\ref{Eq. q_beta}). \footnote{For completeness, \ref{App. linear stability} presents a linear stability analysis of the quiescent state, characterizing the primary phase transitions directly from the spectrum of the synaptic matrix $\mathcal{W}$.}
\begin{figure*}[h!]
    \centering
\includegraphics[width=0.9 \linewidth]{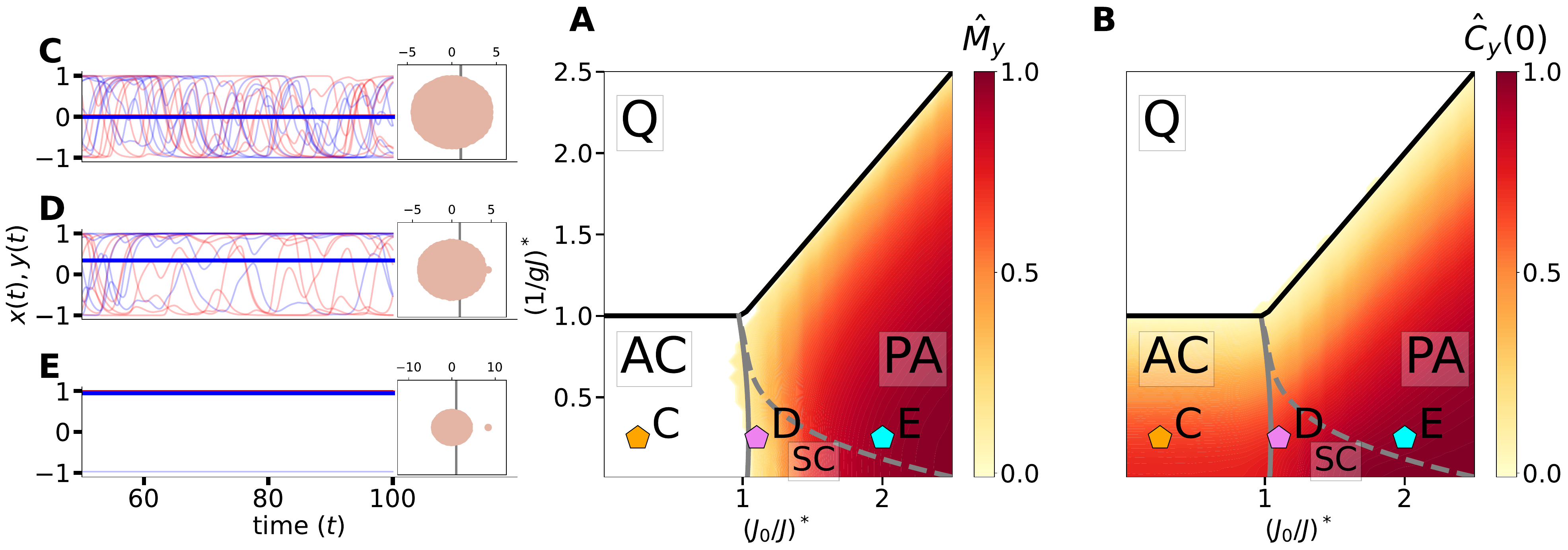}
\includegraphics[width=0.9 \linewidth]{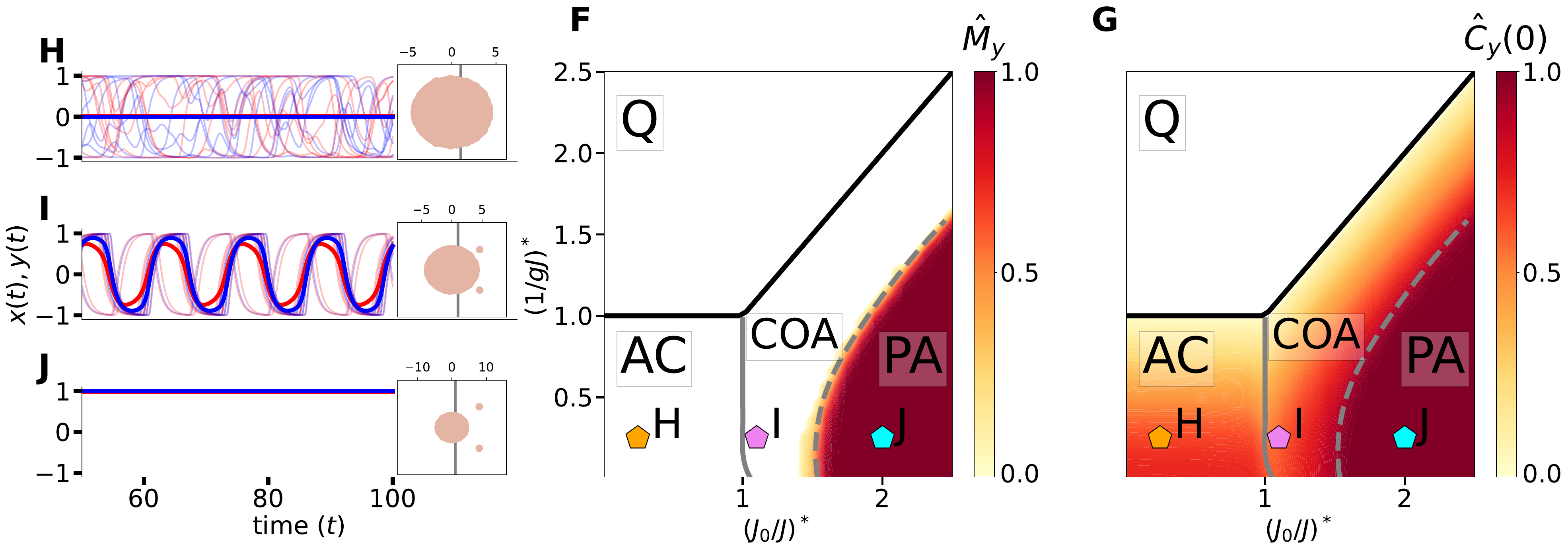}
\includegraphics[width=0.9\linewidth]{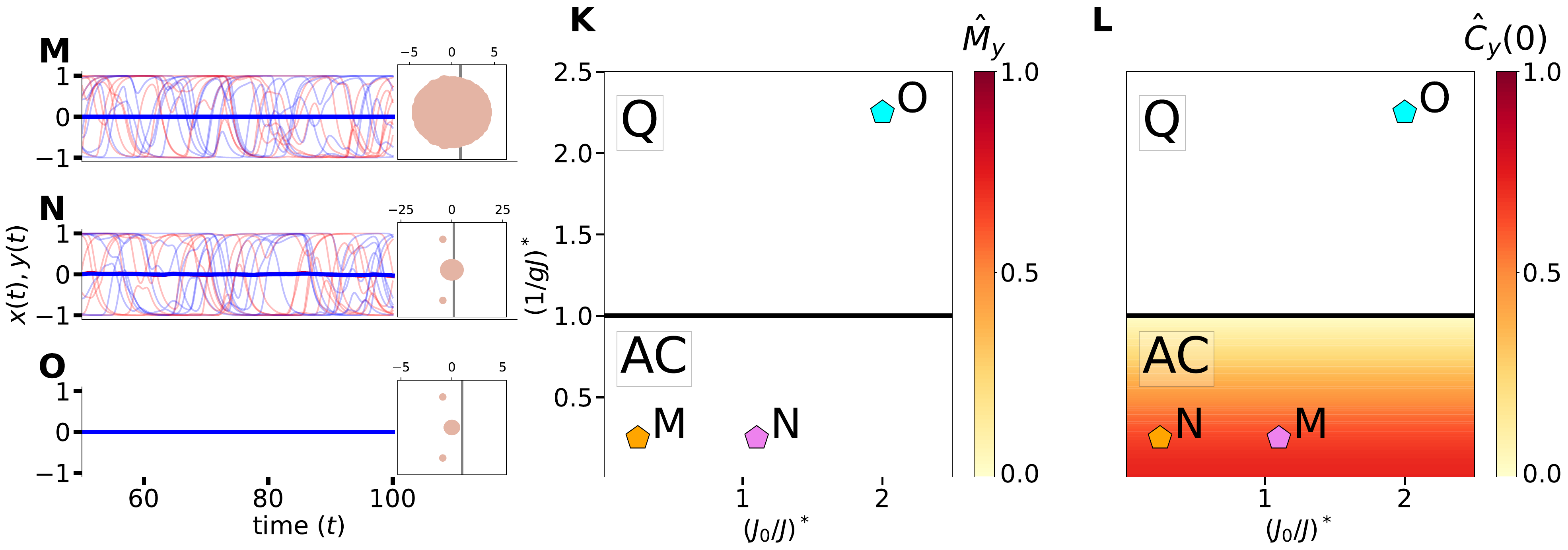}
    \caption{\textbf{Phase diagrams and representative trajectories for $(\beta,\delta)=(0.50,0.50)$} (point A in Fig.~1; \textbf{panels A–E}),\textbf{ $(\beta,\delta)=(0.50,0.25)$ }(point B in Fig.~1; \textbf{panels F–J}), \textbf{ and $(\beta,\delta)=(1.50, 1.25)$ }(point C in Fig.~1; \textbf{panels K–O}). Panels A,B,F,G, K, L show heat maps of the time-averaged mean inhibitory activity $\hat{M}_y$ and the equal-time autocorrelation $\hat{C}_y(0)$. Solid black lines are bifurcation curves from fixed-point stability [Eqs.~(\ref{Eq. fp1 beta},\ref{Eq. fp2 beta})], while the dashed and solid gray lines are transitions of stationary chaotic/oscillatory states from Eqs.~(\ref{Eq. chaos_beta 2}) and (\ref{Eq. chaos_beta 1}), respectively. Panels C–E (for $(\beta,\delta)=(0.50,0.50)$),  H–J (for $(\beta,\delta)=(0.50,0.25)$) and  M–O (for $(\beta,\delta)=(1.50,1.25)$) display trajectories in the Q, AC, and PA or COA phases, with excitatory (red) and inhibitory (blue) population activities; mean activities are highlighted with thicker lines. These examples correspond to parameter points marked by orange, pink, and cyan pentagons in panels A,B,F,G, K and L. Insets in the first column show the eigenvalue spectrum  of the synaptic matrix $\mathcal{W}$ in the complex plane (vertical, gray line indicates $\mbox{Re}(\lambda) = 1$).}
    \label{fig: phase_space}
\end{figure*}

First, we analyse bifurcations in the means $M_x$ and $M_y$. Linearizing Eq. (\ref{Eq. M_beta}) around the quiescent solution gives the instability condition
\begin{equation} \label{Eq. fp2 beta}
\frac{J_0}{J} > \frac{1}{\mathrm{Re}\big(\lambda_M(\mathcal{N}_{\beta, \delta})\big)} \; \frac{1}{gJ},
\end{equation}
where
\begin{equation} \label{Eq. lambda_M N}
\lambda_M(\mathcal{N}_{\beta, \delta}) = \frac{1}{2} \left( 1 - \delta + \sqrt{(1 + \delta)^2 - 4\beta} \right)
\end{equation}
is the eigenvalue of $\mathcal{N}_{\beta, \delta}$ with largest real part. 

The value of $\lambda_M$ is uniquely determined by the target-specific inhibitory parameters $(\beta,\delta)$. 
Therefore, the nature of the emergent phase diagram can be classified directly in terms of $(\beta,\delta)$, as summarized in Fig. \ref{fig:1}.

In terms of $\lambda_M$, Eq.~(\ref{Eq. fp2 beta}) leads to distinct phase transitions (that we refer to as type I transitions). 
When $\mathrm{Re}\big(\lambda_M(\mathcal{N}_{\beta,\delta})\big) > 0$, crossing the instability line in Eq.~(\ref{Eq. fp2 beta}) yields either: 
a transition to a persistent-activity (PA) fixed point, if $\lambda_M$ is real and positive (which corresponds to condition S1 in Fig. \ref{fig:1}); or a transition to coherent oscillatory activity (COA), if $\lambda_M$ is complex with positive real part (condition S2). 
In contrast, when $\mathrm{Re}\big(\lambda_M(\mathcal{N}_{\beta,\delta})\big) < 0$ (condition S3), all eigenvalues of $\mathcal{N}_{\beta,\delta}$ have negative real part and the mean activity cannot bifurcate. \footnote{In the following, we mainly focus on values of $(\beta,\delta)$ leading to a mean-driven bifurcation, namely conditions S1 and S2 in Fig~\ref{fig:1}. Condition S3, however, is also relevant, as it describes states (e.g.\ point C in Fig.~\ref{fig:1}) for which no mean-driven bifurcation occurs. In these cases, the mean inputs vanish, $\mu_x=\mu_y=0$, and the state is therefore \textit{balanced} in the sense defined above. This shows that the strictly balanced case $(\beta=\delta=1)$ is sufficient, but not necessary, for balance --which can also arise under inhibition-dominated conditions such as $1 \leq \delta \leq \beta$.}

Next, we examine bifurcations in $q_x$ and $q_y$ at fixed $M_x = M_y = 0$ (which define type II transitions). Linearization of  Eq.~(\ref{Eq. q_beta}) around the quiescent solution yields a second instability condition
\begin{equation} 
\label{Eq. fp1 beta}
\frac{1}{gJ} \frac{1}{\sqrt{\lambda_M(\mathcal{M}_{\beta, \delta})}} < 1,
\end{equation}
where
\begin{equation} \label{Eq. lambda_M M}
\lambda_M(\mathcal{M}_{\beta, \delta}) = \frac{1}{2} \left( (1 + \delta^2) + \sqrt{(1 - \delta^2)^2 + 4\beta^2} \right)
\end{equation}
is the largest eigenvalue of $\mathcal{M}_{\beta, \delta}$. When Eq.~(\ref{Eq. fp1 beta}) holds, the quiescent state loses stability and a zero-mean, fluctuation-dominated regime --\emph{asynchronous chaos}-- emerges. \footnote{The label \emph{asynchronous} is widely used (e.g., \cite{harish_asynchronous_2015,dahmen_second_2019,li_tuning_2020}) for the chaotic, weakly correlated regime of firing-rate models, reflecting the irregular, fluctuation-driven cortical activity commonly referred to as the \emph{asynchronous state} \cite{renart_asynchronous_2010}.}

In summary, fixed-point stability analyses identify two distinct routes out of the quiescent state, that we called type I and type II, governed by Eq. (\ref{Eq. fp2 beta}) and Eq. (\ref{Eq. fp1 beta}), respectively.

Before proceeding, let us remark that these results suggest a natural rescaling,
\begin{eqnarray}
\left(\frac{J_0}{J}\right)^* &=& 
\dfrac{\Big \vert \text{Re}\big(\lambda_M(\mathcal{N}_{\beta, \delta})\big)\Big \vert }
{\sqrt{\text{Re}\big(\lambda_M(\mathcal{M}_{\beta, \delta})\big)}} \, \frac{J_0}{J}.
\\[4pt]
\left(\frac{1}{gJ}\right)^* &=& 
\dfrac{1}{\sqrt{\text{Re}\big(\lambda_M(\mathcal{M}_{\beta, \delta})\big)}} \, \frac{1}{gJ}, 
\end{eqnarray}
This rescaling provides a unified framework for comparing phase diagram across different $(\beta,\delta)$: after a simple axis redefinition, the quiescent-state stability boundaries in the $(J_0/J,1/gJ)$ plane collapse onto a common curve. In contrast, transitions between non-quiescent states depend explicitly on $\beta$ and $\delta$, so the full phase diagrams cannot be made equivalent by this rescaling.

\subsection{Numerical results}
To complement the DMFT analysis and validate the preceding theoretical predictions, we performed extensive numerical simulations and report here representative results for three parameter sets: $(\beta,\delta)=(0.5,0.5)$, satisfying condition S1; $(\beta,\delta)=(0.5,0.25)$, satisfying condition S2; and $(\beta,\delta)=(1.5,1.25)$, satisfying condition S3. These cases correspond, respectively, to points A, B, and C in Fig.~\ref{fig:1}, left panel.
\footnote{In the Supplemental Material \cite{martorell_supplemental_nodate}, we further examine two additional parameter configurations, $(\beta,\delta)=(0.25,0.50)$ and $(1.00,0.50)$.}
The phase diagrams in Fig.~\ref{fig: phase_space} confirm the transitions predicted by the fixed-point stability analysis and reveal the emergence of three qualitatively distinct types depending on the values of the two inhibitory couplings    .

We numerically compute the time-averaged mean inhibitory activity $\hat{M}_y$ (panels A, F and K) and the equal-time autocorrelation of inhibitory activity $\hat{C}_y(0)$ (panels B, G, L), defined as
\begin{eqnarray}
    \hat{M}_y &=& \frac{1}{T}\int_{t_0}^{t_0 + T} \hat{M}_y(s) \,  ds,  \\  \hat{C}_y(\tau) &=& \frac{1}{T}\int_{t_0}^{t_0 + T} \hat{C}_y(s, s + \tau) \, ds.
\end{eqnarray}
where $\hat{M}_y(t)$ and $\hat{C}_y(t, s)$ are respectively defined by Eqs. (\ref{Eq. hat M})-(\ref{Eq. hat C}). Both quantities are averaged over $S = 1000$ realizations of a system of size $N = 1000$, and time interval $T = 1000$ time-units. The heat maps in Fig. \ref{fig: phase_space}, panels A, B, F, G, K and L, depict both observables across the rescaled phase diagram $\big( (J_0/J)^*, (1/gJ)^*\big)$. 

The phase transitions obtained from fixed-point analysis, Eqs. (\ref{Eq. fp2 beta}) and (\ref{Eq. fp1 beta}), are indicated by solid black lines, while dashed and solid gray lines represent transitions between chaotic and oscillatory regimes, identified through non-fixed-point analysis (see Sec. \ref{Sec. Non-fixed points}).

For $(\beta, \delta) = (0.5, 0.5)$ (point A in Fig. \ref{fig:1}; panels A–B in Fig.~\ref{fig: phase_space}), the phase diagram is divided into three primary phases: Quiescent (Q), Asynchronous Chaos (AC), and Persistent Activity (PA), as determined by fixed-point analysis. Panels C, D, and E illustrate representative microscopic trajectories for each phase, corresponding to the orange, violet, and blue pentagons in panels A and B. These panels display the dynamics of excitatory (red) and inhibitory (blue) neurons, $x_i(t)$ and $y_i(t)$, along with their mean activities $\hat{M}_{x,y}(t)$ (thicker lines). The spectral distribution of the synaptic matrix $\mathcal{W}$ is also shown for each case. In particular, panel D suggests the emergence of an intermediate phase ---named Synchronous Chaotic state (SC)---,
 characterized by chaotic activity with a non-zero mean, similar to previous observations in the one-population model  \cite{mastrogiuseppe_linking_2018, martorell_ergodicity_2025}. 

For $(\beta, \delta) = (0.5, 0.25)$ (point B in Fig. \ref{fig:1}; panels F–G in Fig. \ref{fig: phase_space}), the phase diagram is characterized by Q, AC, and Coherent Oscillatory Activity (COA) phases. Panels H, I, and J depict representative trajectories corresponding to the orange, violet, and blue pentagons in panels F and G. 

For $(\beta,\delta)=(1.50, 1.25)$ (point C in Fig.~\ref{fig:1}; panels K–O in Fig.~\ref{fig: phase_space}), the phase diagram is split into Q and AC phases, as predicted. Accordingly, the only phase transition is the one given by Eq. (\ref{Eq. fp1 beta}) (black line in panel K and L). Panels M and N show representative trajectories of AC states (orange and violet pentagons in panels F and G, respectively), while panel O shows the dynamics of the Q state (blue pentagon).

In summary, our analysis shows that type I transitions from the Q phase lead either to PA or to COA, depending on $(\beta,\delta)$, whereas type II transitions give rise to AC. 
In the particular case of conditions S1 and S2, numerical simulations further show that the nature of the intermediate regime depends on $(\beta,\delta)$: the system either develops a synchronous-chaotic (SC) state between AC and PA, or exhibits a coherent oscillatory (COA) phase. 
The strong agreement between simulations and analytical predictions validates the derived stability criteria and motivates a systematic DMFT analysis of non-fixed-point solutions and their stability, with the goal of fully characterizing the phase diagram.

\subsection{Non-Fixed-Point Stationary Solutions}
\label{Sec. Non-fixed points}

Here we extend the DMFT analysis to non–fixed-point stationary states, defined by time-translation invariance of the dynamics.\footnote{Note that oscillatory (periodic) dynamics is not stationary in this sense.} 
In these regimes, the mean activity is time-independent, $\mathbf{M} = [M_x,M_y]^T$, and the autocorrelation $\mathbf{C}(t,s) = [C_x(t,s),C_y(t,s)]^T$ depends only on the time difference $\tau = t-s$. 
In the particular case of fixed-point dynamics, the autocorrelation reduces to $\mathbf{q} = [q_x,q_y]^T$, defined by Eq. (\ref{Eq. q_beta}).
Such states have been extensively studied in one-population models \cite{schuecker_optimal_2018,crisanti_path_2018,martorell_dynamically_2024,martorell_ergodicity_2025} and in multi-population settings \cite{kadmon_transition_2015, mastrogiuseppe_linking_2018}.

Technical details of the DMFT solutions and their linear-stability analysis are deferred to \ref{App. stationary dynamics} and \ref{App: ls beta}, respectively; here we only summarize the main results. As for fixed points, also in this case there are two distinct routes to instability: one driven by destabilization of the mean activity and another by destabilization of the autocorrelation function.
 The corresponding stability criteria are
\begin{eqnarray}
    \label{Eq. chaos_beta 2}
N_1 \mathrm{Re}\left( 1 + \sqrt{1 - 4 (\beta - \delta ) N_2}\right) &<& \frac{1}{g J_0},\\
\label{Eq. chaos_beta 1}
M_1 \mathrm{Re}\left( 1 + \sqrt{1 + 4 (\beta^2 - \delta^2 ) M_2}\right) &<& \frac{1}{(gJ)^2},
\end{eqnarray} 
respectively, where \(N_1, N_2, M_1,\) and \(M_2\) are numerical coefficients depending on \(\delta\) and on averages of the nonlinear input, \(\langle \phi'(z_{x,y}) \rangle\). Their explicit expressions are given in \ref{App: ls beta}, Eq.~(\ref{Eq. N1 N2}) and Eq. (\ref{Eq. M1 M2}). 

The derived stability conditions enable the identification of phase boundaries between non-quiescent states, as summarized below (we refer to Fig. \ref{fig: phase_space}). 
\begin{itemize}
    \item \textbf{Transitions from the PA phase}: The transition from the PA phase is determined through a linear stability analysis of fixed-point solutions, where \( \bM \neq 0 \) and \( \bq \neq 0 \) (defined by Eqs. (\ref{Eq. M_beta}) and (\ref{Eq. q_beta})). Depending on the parameters $(\beta, \delta)$, two different situations may appear.
     
    \emph{Transitions to SC}
    (under the condition S1 in Fig~\ref{fig:1}).  Due to the nature of SC dynamics, this transition must occur via an instability of the fixed-point autocorrelation, \( \bq \). 
    The stability criterion is therefore given by Eq. (\ref{Eq. chaos_beta 1}), where \( \bC(\tau) = \bq \) for all values of \( \tau \). As a function of the rescaled parameters \( ((J_0/J)^*, (1/gJ)^*) \), Eq. (\ref{Eq. chaos_beta 1}) defines the phase transition indicated as a dashed, gray line in Fig. \ref{fig: phase_space}, panels A and B.  
    
    In the particular case where \( \beta =  \delta\), the stability condition coincides exactly with the one derived for the one-population SCS model, as reported in \cite{mastrogiuseppe_linking_2018, martorell_ergodicity_2025}. 

        \emph{Transitions to the COA regime} (under the condition S2 in Fig. \ref{fig:1}). 
        The PA phase loses stability when the mean activity $\bM$ becomes unstable and starts to oscillate, while the autocorrelation remains constant, $\bC(\tau)=\bq$. 
        The corresponding stability criterion is given by Eq.~(\ref{Eq. chaos_beta 2}) and defines the PA-COA transition, shown as the dashed gray line in Fig.~\ref{fig: phase_space},  panels F and G.

    \item \textbf{Transitions from the AC phase}:
    The transition from AC to either SC or COA is controlled by a bifurcation of the mean activity from chaotic solutions with $\bM = 0$, as given by Eq.~(\ref{Eq. chaos_beta 2}). In the AC phase, the stationary state is constrained to $\bM=0$, and the zero-time autocorrelations $\bC(0)$ are those of the chaotic attractor dynamically selected by the system \cite{martorell_dynamically_2024}; their values must be obtained numerically from the DMFT equations, as described in \ref{App: numerical DMF}.  
   The resulting condition yields the bifurcation line separating AC and SC (solid gray line in Fig.~\ref{fig: phase_space}, panels A and B); or, alternatively, the AC–COA transition (solid gray line in Fig.~\ref{fig: phase_space}, panels F and G). In the symmetric case, $\beta=\delta$, it reduces to the equivalent stability condition of the one-population SCS model \cite{mastrogiuseppe_linking_2018,martorell_ergodicity_2025}.
    \end{itemize}

\begin{figure*}[htb]
    \centering
\includegraphics[width=0.85
\linewidth]{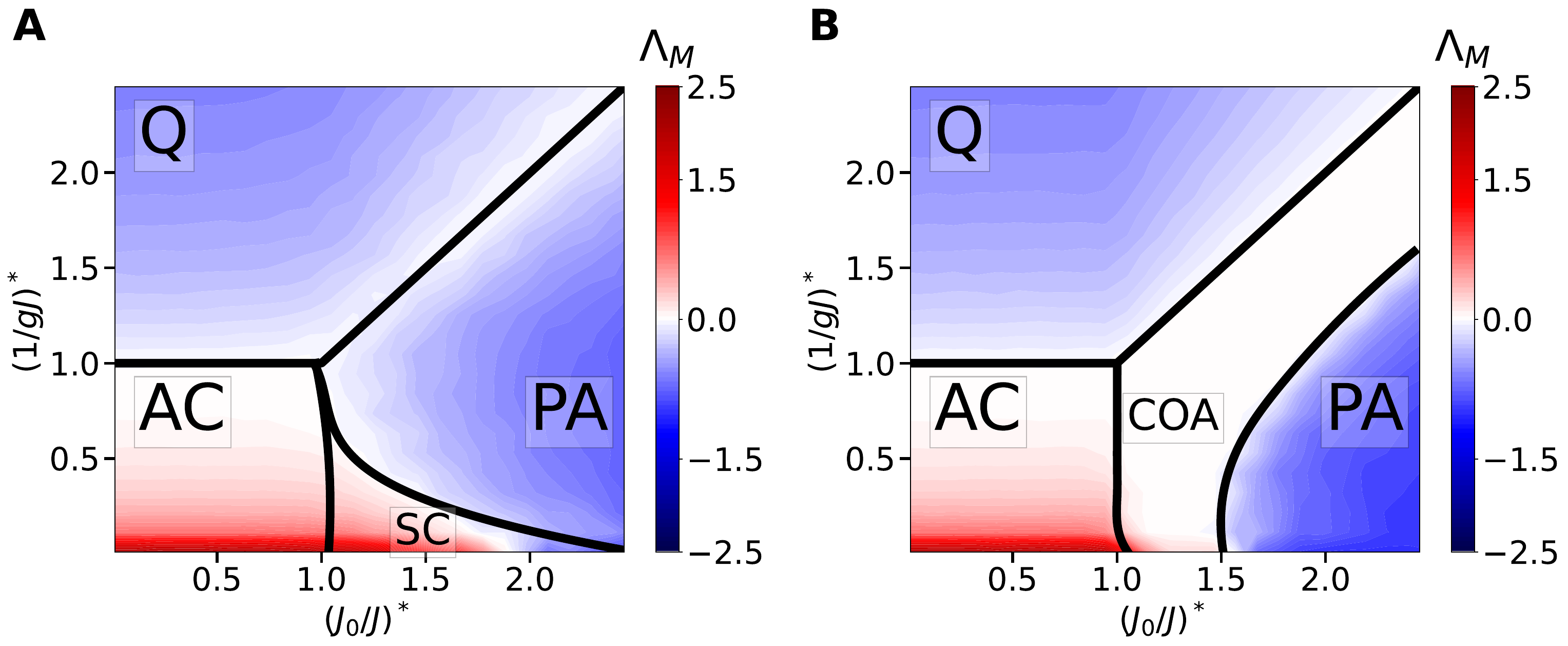}
\caption{\textbf{Largest Lyapunov exponent across dynamical regimes.} Heat map of the largest Lyapunov exponent (LLE) $\Lambda$ for a network of size $N = 1000$ with $(\beta, \delta) = (0.5, 0.5)$ (\textbf{panel A}) and $(\beta, \delta) = (0.5, 0.25)$ (\textbf{panel B})  plotted as a function of the rescaled parameters $((J_0/J)^*, (1/gJ)^*)$. Positive values of $\Lambda$ identify chaotic regimes, while oscillatory activity is linked with a zero value. The LLE has been averaged over $S = 100$ different realizations. Bifurcation lines are depicted identically to those of Fig. \ref{fig: phase_space}.}
    \label{fig:3}
\end{figure*}

 In summary, the stability analysis of stationary solutions provides a comprehensive framework for characterizing non-quiescent phase transitions in the E/I firing-rate model. Unlike pure fixed-point analysis, which misses some dynamical transitions, it reveals structured non–fixed-point stationary states. Taken together, the fixed-point and stationary-state analyses yield a unified description of the phase diagram (as depicted in Fig.~\ref{fig: phase_space}), with three principal regimes --Quiescent (Q), Asynchronous Chaos (AC), and Persistent Activity (PA)-- and two intermediate phases, Synchronous Chaos (SC) and Coherent Oscillatory Activity (COA). Crucially, the phase diagrams fall into three distinct classes: those in which the non-quiescent transition destabilizes to a SC, those in which it destabilizes to a COA and those that display only Q--AC transitions.

\section{Emergent Phases Beyond the Asynchronous Chaos}

In this section, we investigate the dynamical nature of the intermediate phases emerging in the E/I network, in particular the SC and COA phases. We therefore restrict the analysis to values of $(\beta,\delta)$ satisfying conditions S1 and S2 in Fig. \ref{fig:1}.

\subsection{Characterization of Chaos}

To quantify numerically the chaotic behaviour of the network, we compute the largest Lyapunov exponent (LLE), denoted by $\Lambda_M$, from simulations of the microscopic dynamics described by Eq.~(\ref{eq:dynamics-def}). The LLE is estimated using the standard Benettin--Wolf algorithm \cite{benettin_lyapunov_1980, wolf_determining_1985, pikovsky_lyapunov_2016}. Fig. \ref{fig:3} shows $\Lambda_M$ for a network of size $N = 1000$ with $(\beta, \delta) = (0.5, 0.5)$ (panel A, corresponding to point A in Fig. \ref{fig:1}) and $(\beta, \delta) = (0.5, 0.25)$ (panel B, corresponding to point B in Fig. \ref{fig:1}), plotted as a function of the rescaled parameters $((J_0/J)^*, (1/gJ)^*)$. 

In Fig.~\ref{fig:3}, panel A, chaotic activity arises in the AC phase and extends into the SC phase, closely resembling the phase diagram of the one-population SCS model \cite{martorell_ergodicity_2025}, with the boundary of the chaotic region —characterized by positive LLEs— coinciding with the analytical phase boundaries (Q–AC and PA–SC transitions) derived using DMFT techniques. In contrast, in panel B, chaos arises only within the AC phase and transitions to an oscillatory phase (AC–COA) with vanishing LLE, consistent with \cite{rajan_stimulus-dependent_2010}, where a sufficiently strong, coherent oscillatory drive --here endogenously generated-- suppresses asynchronous chaos; while for sufficiently large $(J_0/J)^*$ the system instead reaches the PA phase, where perturbations decay and the dynamics are Lyapunov-stable (negative LLE).
\footnote{The value $\Lambda_M=0$ observed in the COA phase in Fig.~\ref{fig:3}, panel B, is consistent with its oscillatory character. Autonomous continuous-time dynamics with stable periodic motion are Lyapunov-marginally stable along the direction of the oscillation \cite{strogatz_nonlinear_2019, guckenheimer_nonlinear_2002}.}

\subsection{Characterization of Oscillations and Synchronization}

\begin{figure*}[t]
    \centering
\includegraphics[width=1
\linewidth]{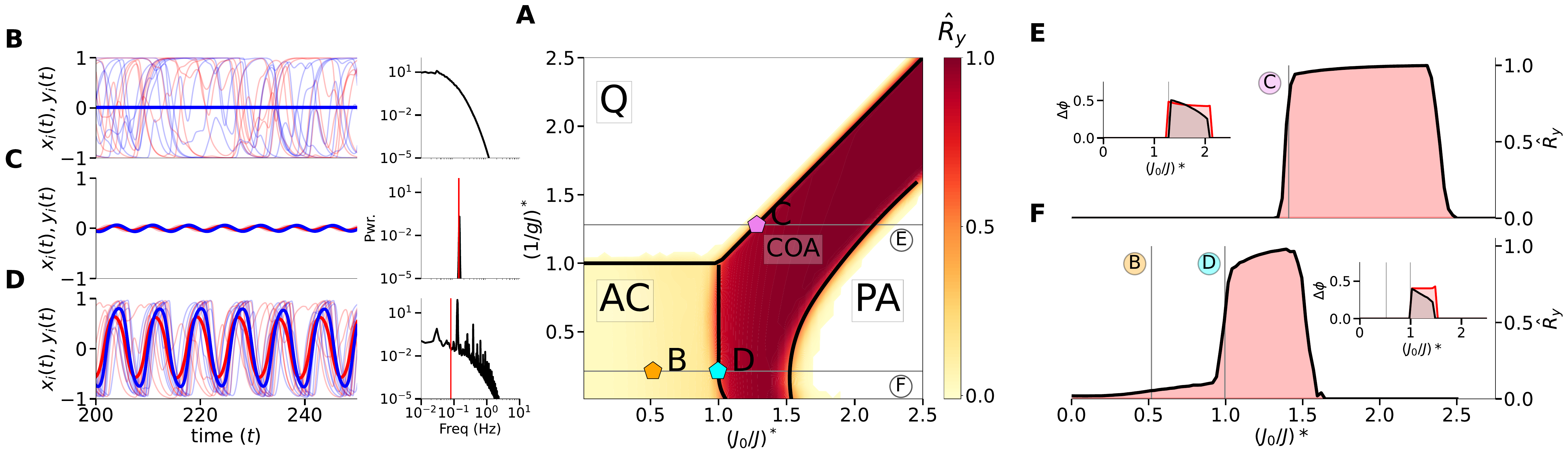}
\caption{
\textbf{Oscillatory regimes and transition routes to the COA phase.} 
\textbf{Panel A}: Time-averaged Kuramoto parameter $R_y$ of the inhibitory population as a function of the rescaled parameters $((J_0/J)^*, (1/gJ)^*)$, indicating the degree of global phase synchronization. 
\textbf{Panels B--D}: Representative dynamical states corresponding to colored pentagons in panel A. 
These panels show the excitatory (red) and inhibitory (blue) activity for some neurons, $x_i(t)$ and $y_i(t)$, and the mean activities $\hat{M}_{x,y}(t)$ (thicker lines),
together with the population-averaged power spectrum within the AC phase, at $((J_0/J)^*, (1/g J)^*) = (0.25, 0.25)$; and COA phase, at $(1.25, 1.25)$ and $(1, 0.25)$, respectively.
\textbf{Panels E--F}: Kuramoto parameter $R_y$ as a function of $(J_0/J)^*$, for $(1/g J)^* = 1.25$ (panel E) and $(1/g J)^* = 0.25$ (panel F), as indicated in panel A. The insets show the corresponding phase difference $\Delta \phi$ computed from simulations (black line) and compared with the theoretical prediction (red line), Eq.(\ref{Eq. theoretical Delta phi}). Numerical results correspond to simulations of the microscopic system Eq.(\ref{eq:dynamics-def}), for $N = 2000$, averaged over $S = 200$ different realizations and $(\beta, \delta) = (0.5, 0.25)$. 
} \label{fig:4}
\end{figure*}

In the previous section, within the DMFT framework, we indicated that oscillatory behaviour arises when the mean activity vector $\mathbf{M}(t)$ becomes unstable through a dominant complex eigenmode of the Jacobian matrix governing its linearized dynamics, Eq.~(\ref{Eq. JMM}). Technical details of this derivation are provided in \ref{App. Dynamics Mean}.

For $(\beta,\delta)$ in the parameter region specified by condition S2 in Fig. \ref{fig:1}, this Jacobian matrix admits a pair of complex-conjugate eigenvalues. We denote the corresponding eigenvectors and eigenvalues by
\begin{equation}
    \mathbf{v}_M^{\pm} = \big[e^{i \phi^\pm_x},\, e^{i \phi^\pm_y}\big]^T, \qquad 
    \lambda^{\pm}_M = r \pm i\, \omega_0,
\end{equation}
where $i$ denotes the imaginary unit. Close to the AC-COA transition, the mean activity is therefore dominated by the oscillatory mode, such that
\begin{equation}
    \mathbf{M}(t) \sim \mathbf{v}_M^{\pm}\, e^{i \omega_0 t},
\end{equation}
with oscillation frequency
\begin{equation} \label{Eq. EI omega_0}
    \omega_0 = g J_0 N_1 \sqrt{4(\beta - \delta) N_2 - 1}.
\end{equation}
The phase lag between the excitatory and inhibitory mean activities, $M_x(t)$ and $M_y(t)$, is set by the phase difference between the components of the dominant complex eigenvector, 
\begin{equation} \label{Eq. theoretical Delta phi}
    \Delta \phi = \phi^{\pm}_x - \phi^{\pm}_y
    = \arctan \!\left(
    \kappa\, \frac{\sqrt{4(\beta - \delta) N_2 - 1}}{1 - \kappa}
    \right),
\end{equation}
where $\kappa = N_1 / \langle \phi'(z_x) \rangle$. The function $\arctan(x)$ is defined as the principal branch of the inverse tangent, so that the phase lag satisfies $\Delta \phi \in (-\pi/2, \pi/2)$.

To characterize the oscillatory dynamics beyond linear theory, we simulated the microscopic dynamics and computed, for each population, a Kuramoto order parameter from the instantaneous phases obtained via the Hilbert transform \cite{cohen_time_1995}. This quantity measures the degree of phase coordination within each population; see \ref{App. Kuramoto} for details on the numerical procedure. Results for $(\beta, \delta) = (0.5, 0.25)$ are reported in Fig.~\ref{fig:4}. 

In Fig.~\ref{fig:4}, panel A displays the time-averaged Kuramoto parameter $R_y$ of the inhibitory population as a function of the rescaled parameters $((J_0/J)^*, (1/gJ)^*)$.
In Fig.~\ref{fig:4}, panels E--F show the parameter $R_y$ for two different sections of the phase diagram, $(1/g J)^* = 1.25$ (panel E) and  $(1/g J)^* = 0.25$ (panel F), as indicated in panel A. Each inset shows the corresponding phase difference $\Delta \phi$ between the mean activities $\hat{M}_x(t)$ and $\hat{M}_y(t)$ computed from simulations (black line) and compared with the theoretical prediction (red line), Eq.  (\ref{Eq.  theoretical Delta phi}). 
In Fig.~\ref{fig:4}, panels B--D illustrate representative dynamical regimes, where we plot the excitatory (red) and inhibitory (blue) activity for some neurons, $x_i(t)$ and $y_i(t)$, and the mean activities $\hat{M}_{x,y}(t)$ (thicker lines); 
together with the population-averaged power spectrum of the activity.

In the AC phase (Fig.~\ref{fig:4}, panel A), the mean activity is weak and no global coordination emerges (Fig.~\ref{fig:4}, panel B for representative AC dynamics): due to the finite size of the system a residual synchronization appears, leading to a small and fluctuating Kuramoto order parameter. The power spectrum exhibits a broad, slowly decaying profile without a dominant peak. 

Within the COA phase the Kuramoto parameters have a finite value (Fig.~\ref{fig:4}, panel A), indicating a high degree of phase synchronization. In this regime, the oscillatory mean activity generates a strong effective E/I feedback loop, reflected in a non-zero phase lag $\Delta \phi > 0$ between excitatory and inhibitory populations (see insets in Fig.~\ref{fig:4}, panels E and F). The COA phase can be reached via two distinct routes: through a transition from the quiescent (Q) phase (Fig. ~\ref{fig:4}, panel E) or from the AC phase (Fig. ~\ref{fig:4}, panel F), each associated with qualitatively different oscillatory dynamics.

The Q--COA transition occurs when the entire network starts oscillating coherently in the mean-field mode, $\mathbf{v}^{\pm}_M e^{i \omega t}$. In this case, quenched disorder plays a minor role. Accordingly, the power spectrum displays a sharp peak at $\omega_0$ (Fig.~\ref{fig:4}, panel C, vertical red line), and the Kuramoto parameter exhibits a rapid increase, signalling an abrupt or explosive onset of global synchronization (Fig.~\ref{fig:4}, panel E).

In contrast, at the AC--COA transition the emergent oscillations compete with the pre-existing asynchronous chaotic activity. The resulting dynamics reflects a non-trivial interplay between oscillatory order and chaotic fluctuations: near the transition, the power spectrum retains a broad, slowly decaying background characteristic of chaos, on top of which a peak appears at the oscillation frequency $\omega_0$ (Fig.~\ref{fig:4}, panel D). This chaos-to-oscillation route is reminiscent of the stimulus-induced suppression of chaos reported in \cite{rajan_stimulus-dependent_2010, fournier_high-dimensional_2025}, although here the oscillatory drive is generated endogenously by the mean activity. Because of residual disorder and the underlying chaotic component, the system is only partially synchronized at the transition, and the Kuramoto parameters remain below the fully coherent value (Fig.~\ref{fig:4}, panel F). The synchronization gradually increases as the system is driven deeper into the COA phase (for instance, $R_y$ increases from 0.9 to 1.0 in panel F within the COA phase).
\footnote{As discussed above, no genuine synchronization is expected within the AC phase, since $M_x$ and $M_y$ vanish. The small but positive Kuramoto parameter observed (see, e.g., Fig.~\ref{fig:4}, panel A) is therefore a finite-size effect. This residual apparent synchronization becomes more pronounced close to the AC–COA transition as a consequence of critical slowing down near the bifurcation (see, e.g., Fig.~\ref{fig:4}, panel F) .}

\section{Discussion and Conclusions}

Cortical circuits can exhibit a rich repertoire of dynamical regimes—from asynchronous irregular activity to persistent states and coherent oscillations—depending primarily on how excitation, inhibition, and connectivity structure interact. Understanding how these regimes are organized, and what mechanisms control the transitions between them, is therefore a central problem at the interface of statistical physics and theoretical neuroscience. For example, the \emph{criticality hypothesis} proposes that cortical networks may operate near phase transitions, where activity can display scale invariance and favorable computational properties such as enhanced dynamic range and information transmission \cite{beggs_neuronal_2003,beggs_criticality_2008,shew_functional_2013,mora_are_2011,munoz_colloquium_2018,dearcangelis_self-organized_2006,deco_dynamics_2017,wilting_25_2019,obyrne_how_2022,hengen_is_2025,cocchi_criticality_2017}. From this perspective, a key theoretical question is to understand how the different dynamical regimes (or phases) are organized in parameter space and what transitions separate them.

A major step in this direction was provided by the  seminal work of Sompolinsky--Crisanti--Sommers (SCS) framework \cite{sompolinsky_chaos_1988}, which uncovered the onset of high-dimensional chaos in randomly connected networks and established a direct link between microscopic connectivity statistics and macroscopic dynamics in the thermodynamic limit \cite{kadmon_efficient_2025}. Building on this framework, a large body of work has shown that adding low-rank structure to otherwise random connectivity introduces a small number of low-dimensional collective modes that interact with disorder-driven fluctuations and thereby reshape the dynamical phase diagram \cite{mastrogiuseppe_linking_2018,schuessler_dynamics_2020,beiran_shaping_2021,dubreuil_role_2022}. Within this general setting, such collective modes can stabilize non-trivial fixed points, generate coherent oscillations, and organize broken-symmetry chaotic states on top of an otherwise high-dimensional fluctuating background \cite{schuecker_optimal_2018,kadmon_transition_2015,martorell_dynamically_2024,martorell_ergodicity_2025,landau_coherent_2018,garcia_del_molino_synchronization_2013, fournier_high-dimensional_2025, fournier_non-reciprocal_2025}.

Despite this progress, an important limitation remains in the specific case of excitatory--inhibitory (E/I) networks, which naturally induce a rank-two structure in the connectivity. Most previous studies either assume excitation--inhibition balance or impose symmetric effective couplings across populations \cite{mastrogiuseppe_intrinsically-generated_2017,garcia_del_molino_synchronization_2013,harish_asynchronous_2015,kadmon_transition_2015}, thereby restricting the range of dynamical regimes that can emerge. Some works do consider target-specific couplings \cite{mastrogiuseppe_linking_2018, kadmon_transition_2015}, but do not systematically characterize the dynamical nature of the resulting states. In this work, we study a minimal SCS extension of E/I networks with target-specific inhibition, parameterized by $(\beta,\delta)$.

We show within the DMFT framework that target-specific inhibition reorganizes the phase diagram and selects the nature of the dominant collective instability. In particular, we derived a two-population DMFT description and combined it with fixed-point and stationary-state stability analyses. This allowed us to separate disorder-driven instabilities, associated with the random component of the connectivity, from mean-driven instabilities, controlled by its structured component. Within this framework, the  strictly balanced case $(\beta=\delta=1)$ appears as a singular limit: the mean mode is fully suppressed, the model reduces to an effective one-population SCS system, and the only transition is the classical one between quiescent and asynchronous chaos. Away from this limit, the connectivity develops a rank-two structure that has a crucial dynamical impact, generating collective routes towards an oscillatory regime.

The phase diagram is organized by the differential weight  of the two types of 
inhibitory interactions as determined by the  dominant eigenvalue of the effective mean connectivity, $\mathcal{N}_{\beta, \delta}$. When the leading eigenvalue is real and positive, DMFT shows that the system develops an intermediate SC phase between AC and the PA regime. In this case, chaos survives but acquires a non-vanishing collective component, extending to the E/I system the synchronous-chaos phenomenology known from rank-one models \cite{martorell_ergodicity_2025, mastrogiuseppe_linking_2018}. When the dominant eigenvalue is a complex-conjugate pair, DMFT reveals the intermediate phase is oscillatory: the system undergoes a transition from AC to COA before reaching the final PA regime. 
If, by contrast, the dominant eigenvalue has negative real part, the structured mode does not destabilize the mean activity, and the system displays only Q and AC. In this case, the mean inputs vanish and the system is therefore \textit{balanced}.

Thus, for fixed $(\beta,\delta)$, the phase diagram falls into one of three robust classes, depending on whether the dominant eigenvalue is real and positive, forms a complex-conjugate pair with positive real part, or has negative real part.

Importantly, the oscillatory route does not lead to a robust phase of chaos around an oscillatory mean activity. Instead, both DMFT and microscopic simulations indicate that the $\mathrm{AC}\to\mathrm{COA}$ transition reflects a competition between high-dimensional chaotic fluctuations and a low-dimensional oscillatory collective mode. Near the transition, the oscillatory state still carries remnants of the chaotic background together with the emerging coherent rhythm (as also noted in \cite{mastrogiuseppe_linking_2018}), but deeper in the oscillatory phase the collective mode suppresses the chaotic dynamics. This mechanism is reminiscent of stimulus-induced suppression of chaos \cite{rajan_stimulus-dependent_2010, fournier_high-dimensional_2025}, with the important difference that here the suppressing oscillatory drive is not externally imposed but generated endogenously by the structured E/I feedback. 

The excellent agreement between analytical predictions and numerical simulations supports this interpretation. The bifurcation lines obtained from the DMFT stability analysis correctly identify the onset of the different phases, while Lyapunov and Kuramoto analyses clarify their dynamical nature, thereby confirming both the phase boundaries and the qualitative nature of the intermediate regimes.

From a neurobiological perspective, these results confirm and highlight target-specific inhibition as an important organizing principle for cortical dynamics.  Our results suggest that such asymmetries can qualitatively alter the collective state of the network without changing its overall random architecture. This provides a simple mechanistic link between cell-type-specific inhibitory targeting and large-scale dynamical phenomena such as internally generated rhythms, variability suppression, and persistent collective activity.

Several extensions would bring the present framework closer to biological circuits. A first natural step is to consider positively constrained transfer functions, such as rectified nonlinearities, which are known to alter the balance between fluctuating and mean-driven activity \cite{harish_asynchronous_2015,mastrogiuseppe_intrinsically-generated_2017,kadmon_transition_2015}. Other relevant directions include sparse rather than dense connectivity, multiple inhibitory subtypes, structured perturbations beyond rank two, and finite-size effects. These ingredients could modify the extent and robustness of the oscillatory and structured-chaotic regimes, and may help clarify how the present mechanism operates in more realistic cortical settings.

In summary, we have shown that breaking inhibitory symmetry qualitatively reshapes the dynamical landscape of recurrent excitatory--inhibitory networks. A minimal rank-two mean structure is sufficient to generate two distinct classes of phase diagrams, selected by whether the dominant collective mode is associated with a real outlier or with a complex-conjugate pair. In this way, target-specific inhibitory couplings emerge as a structural lever that controls the competition between asynchronous chaos, synchronous chaos, and coherent oscillations in disordered recurrent networks.

\vspace{1cm}
{\bf{Author Contributions:}}
\textbf{Carles Martorell}: Writing – original draft, Validation, Software, Investigation, Formal analysis. \textbf{Rubén Calvo}: Validation, Software, Investigation, Formal analysis. \textbf{Alessia Annibale}: Writing – review \& editing, Writing – original draft, Supervision, Methodology, Formal analysis, Conceptualization. \textbf{Miguel A. Muñoz}: Writing – review \& editing, Supervision, Funding acquisition, Formal analysis, Conceptualization. 

\vspace{1cm}
{\bf{Declaration of competing interest :}} The authors declare that they have no known competing financial interests or personal relationships that could have appeared to influence the work reported in this paper. No experimental data were used. The simulation code is available from the corresponding author upon request.

\vspace{1cm}
{\bf{Acknowledgments:}}
This work has been supported by Grant No. PID2023-149174NB-I00
financed by the Spanish Ministry and Agencia Estatal de Investigaci\'on MICIU/AEI/10.13039/501100011033 and EDRF funds.
We are thankful to Adri Roig and Victor Buendía for useful and inspiring discussions.
%AA acknowledges insightful discussions with Pierfrancesco Urbani and Valentina Ros. 
\vspace{1cm}

\appendix
\section*{APPENDICES}

\section{Random matrix theory considerations}
\label{App. linear stability}
In this appendix, we analyze the stability of the quiescent state $[\bx_0,\by_0]^T=0$ by introducing a linear perturbation of Eq.~(\ref{eq:dynamics-def}), thereby deriving the corresponding stability criterion in terms of the spectral distribution of the synaptic matrix. Writing
\begin{equation}
\bx(t)=\bx_0+\varepsilon\,\delta\bx(t), 
\qquad
\by(t)=\by_0+\varepsilon\,\delta\by(t),
\end{equation}
and expanding to first order in $\varepsilon$ ( \(0<\varepsilon\ll 1\)), Eq.~(\ref{eq:dynamics-def}) yields the linearized system
\begin{equation}
(\partial_t+\mathrm{Id})
\begin{bmatrix}
\delta\bx(t)\\
\delta\by(t)
\end{bmatrix}
=
g\,\mathcal{W}
\begin{bmatrix}
\delta\bx(t)\\
\delta\by(t)
\end{bmatrix},
\end{equation}
where $\mathrm{Id}$ is the identity matrix and $\mathcal{W}$ denotes the Jacobian evaluated at the quiescent fixed point (defined by Eq. (\ref{Eq. W})).
As usual, linear stability requires that all the Jacobian-matrix eigenvalues satisfy \(g \, \mbox{Re}(\lambda) < 1\), ensuring that perturbations decay; eigenvalues with non-zero imaginary part correspond to oscillatory modes.

As each submatrix \( W^{AB} \) is drawn from a Gaussian ensemble, its eigenvalue distribution converges to the Circular (Girko's) Law in the large size limit ($N \to \infty$) \cite{tao_random_2009, tao_outliers_2013, knowles_outliers_2014}: the eigenvalues of \( W^{AB} \) are uniformly distributed within a disk of radius \( J \) in the complex plane, except for a possible outlier at \( J_0 \), with associated  uniform eigenvector $\mathbf{u} = [1, ..., 1]^T$. However, for arbitrary values of \( \beta \) and \( \delta \), the full matrix \( \mathcal{W} \) does not necessarily obey the Circular Law.  

\begin{figure}[htb]
    \centering
\includegraphics[width=1\linewidth]{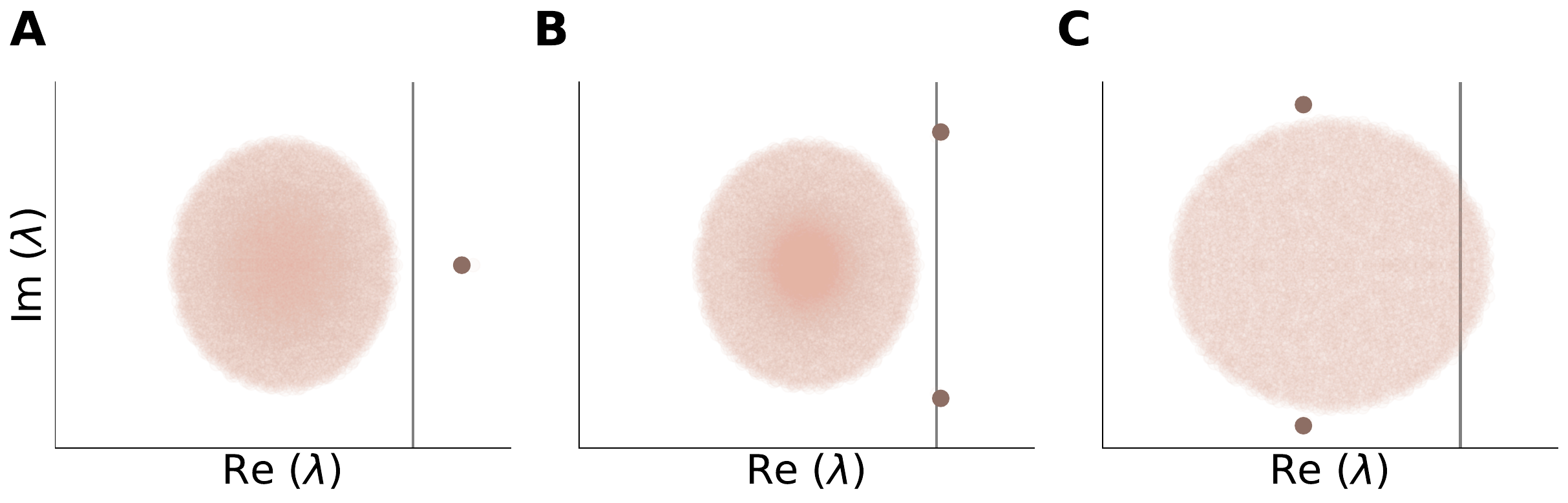}
 \caption{\textbf{Spectral distribution of the synaptic matrix $g\,\mathcal{W}$ in the complex plane.} 
    The bulk spectrum (brown shading, intensity proportional to eigenvalue density) and the possible outliers (dark brown dots) are shown for three representative values of $(\beta,\delta)$: $(0.5,0.5)$ (A), $(0.50,0.25)$ (B), and $(1.40,1.15)$ (C), corresponding to the pentagons A–C in Fig.~\ref{fig:1}. 
    The vertical gray line at $\mathrm{Re}(\lambda)=1$ marks the linear stability threshold. 
    Parameters are fixed to $J_0=0.275$, $J=1$, and $g=2.75$.}
    \label{fig:eig}
\end{figure}

To illustrate deviations from the Circular Law, Fig.~\ref{fig:eig} displays the spectrum of $g\,\mathcal{W}$ for different values of $(\beta,\delta)$, keeping $J_0$, $J$, and $g$ fixed. 
Two robust features are observed: (i) a non-uniform bulk eigenvalue distribution, concentrated around the origin; and (ii) the possible emergence of outliers (dark brown dots), either a single real eigenvalue or a complex-conjugate pair.

The size of the bulk depends on $\beta$ and $\delta$, and in general cannot be analytically characterized (with the notable exception of the $\beta = \delta$ case, as reported by Rajan and Abbott~\cite{rajan_eigenvalue_2006}). In contrast, the possible outliers are easy to identify by noting that their eigenvector is uniform, $ \mathbf{u} = [1,\ldots,1]^T $. Then $[\bx, \by]^T = [\mathbf{u}, \kappa_\pm \mathbf{u}]^T$ are eigenvectors of $\mathcal{W}$, with constant
\begin{equation}
    k_\pm = \frac{1+\delta \mp \sqrt{(1+\delta)^2 - 4\beta}}{2\beta}
\end{equation}
and eigenvalues
\begin{equation}
\label{Eq. outliers}
\lambda_O \equiv \frac{J_0}{2}\left(1-\delta \pm \sqrt{(1+\delta)^2-4\beta}\right).
\end{equation}
These values coincide, up to the prefactor $J_0$, with the leading eigenvalues of the matrix $\mathcal{N}_{\beta,\delta}$, defined in Eq.~(\ref{Eq. lambda_M N}).

Stability is lost when $g\,\mathrm{Re}(\lambda_M)>1$, where $\lambda_M$ denotes the eigenvalue of $\mathcal{W}$ with largest real part. Accordingly, instabilities of the quiescent state fall into two distinct classes \cite{dahmen_second_2019,li_tuning_2020,martorell_dynamically_2024}:
\begin{itemize}
    \item \textit{Type I (outlier-driven)}: an unstable outlier eigenvalue destabilizes the uniform mode $[\mathbf{u},\mathbf{u}]^T$, leading to homogeneous collective activity. If the outlier is real, the transition yields persistent activity (Q-PA); if it forms a complex-conjugate pair, coherent oscillations emerge (Q-COA).
    
    \item \textit{Type II (bulk-driven)}: in the absence of outliers, instability arises from the spectral bulk. In this case, a set of eigenmodes simultaneously approaches the stability threshold, producing fluctuating activity (Q-AC).
\end{itemize}

Type I instabilities are fully characterized within linear stability theory through the analytical expression of the outliers, Eq.~(\ref{Eq. outliers}). In contrast, type II instabilities cannot be captured analytically at this level, as no closed-form expression can be derived for the radius of the spectral bulk for general $(\beta,\delta)$.

\section{Dynamics under Symmetric Inhibition}
\label{App. DMF balance condition}

We consider the symmetric-inhibition condition \(\beta=\delta\equiv b\). In this case, the two rows of the effective coupling matrices, defined by Eqs. (\ref{Eq. N}) and (\ref{Eq. M}), are identical:
\begin{equation}
    N_{b,b}=
    \begin{pmatrix}
        1 & -b\\
        1 & -b
    \end{pmatrix},
    \qquad
    M_{b,b}=
    \begin{pmatrix}
        1 & b^2\\
        1 & b^2
    \end{pmatrix}.
\end{equation}
It follows that the effective mean inputs and noise covariances of the two populations coincide:
\begin{align}
    \mu_x(t)=\mu_y(t)
    &=
    M_x(t)-bM_y(t), \\
    \Delta_x(t,s)=\Delta_y(t,s)
    &=
    C_x(t,s)+b^2C_y(t,s).
\end{align}
Thus, the effective processes \(x(t)\) and \(y(t)\) obey the same stochastic
equation, driven by independent Gaussian fields \(\eta_x(t)\) and
\(\eta_y(t)\) with identical two-time covariances,
$\Delta_x(t,s)$ and $\Delta_y(t,s)$. If their initial conditions are drawn from the
same distribution, the two processes are statistically equivalent. Hence, their
common law can be represented by a single effective process \(z(t)\), defined as
\begin{equation}
    (\partial_t+1)z(t)
    =
    \phi\left[
        gJ_0\mu_z(t)+gJ\eta_z(t)
    \right],
\end{equation}
where \(\eta_z(t)\) is a zero-mean Gaussian process with covariance
\begin{equation}
    \left\langle \eta_z(t)\eta_z(s)\right\rangle
    =
    \Delta_z(t,s).
\end{equation}
The associated order parameters are
\begin{equation}
    M_z(t)=\langle z(t)\rangle,
    \qquad
    C_z(t,s)=\langle z(t)z(s)\rangle,
\end{equation}
and self-consistency yields
\begin{equation}
    \mu_z(t)=(1-b)M_z(t),
    \qquad
    \Delta_z(t,s)=(1+b^2)C_z(t,s).
\end{equation}
Hence, under symmetric target-specific inhibitory couplings, \(x(t)\) and \(y(t)\) are two independent realizations of the same effective process \(z(t)\). In this precise sense, the two-population DMFT system reduces to a one-population DMFT system, with effective mean coupling \((1-b)J_0\) and effective disorder variance \((1+b^2)J^2\).

\section{Formulation of the Stationary Dynamics}
\label{App. stationary dynamics}

For a stationary solution, the mean $\mathbf{M}$ satisfies the following self-consistent equations: 
\begin{equation}
\begin{aligned}
M_x &= \Sigma_x(\bM;\bC(0)) \equiv \int_{-\infty}^{\infty} \mathcal{D}u \, \phi\Big(gJ_0 \, \mu_x + g J\sqrt{\Delta_x}u  \Big), \\
M_y &=\Sigma_y(\bM;\bC(0)) \equiv  \int_{-\infty}^{\infty} \mathcal{D}u\,  \phi\Big(gJ_0 \, \mu_y + g J\sqrt{\Delta_y}u \Big), 
\end{aligned}\label{Eq. M_2_beta}
\end{equation}
where $\mu_{x,y}$ and $\Delta_{x,y}$ are defined in Eqs.~(\ref{Eq. mus}) and (\ref{Eq. deltas}).
The evolution of the autocorrelation function is governed by the following equations, obtained from the DMF formulation in Eq. (\ref{Eq: DMF-beta}): 
\begin{equation}
\begin{aligned} \label{Eq. dynamics C_beta} 
(1-\partial_\tau^2)C_x(\tau) &=& \Xi_x(\mathbf{C}(\tau); \mathbf{C}(0), \mathbf{M}), \\ (1-\partial_\tau^2)C_y(\tau) &=& \Xi_y(\mathbf{C}(\tau); \mathbf{C}(0), \mathbf{M}). 
\end{aligned}
\end{equation}
The functions $\Xi_x$ and $\Xi_y$ denote the lagged-time autocorrelations of the stochastic processes $\phi(z_x(t))$ and $\phi(z_y(t))$, respectively. Here, $z_x(t)$ and $z_y(t)$ represent the instantaneous input to excitatory and inhibitory populations: 
\begin{equation}    \label{Eq: steady-state} 
\begin{aligned}
z_x(t) &=& g J_0 \, \mu_x + g J \, \eta_x(t) , \\
z_y(t) &=& g J_0 \, \mu_y + g J \, \eta_y(t).
\end{aligned}
\end{equation} 

To simplify the notation, these expressions can be replaced by a generic function $\Xi$, defined as the autocorrelation of $\phi(w(t))$, where $w(t)$ is a Gaussian process with mean $\langle w(t) \rangle = g J_0\,  \mu$ and autocorrelation $\langle w(\tau) w(0) \rangle = (g J)^2\,  \Delta(\tau)$. Since the pair $(w(\tau), w(0))$ is jointly Gaussian distributed, the function $\Xi$ can be rewritten in the following integral expression: 
\begin{equation}
\Xi(\Delta; \Delta_0, \mu) = \iiint \mathcal{D}u \mathcal{D}v \mathcal{D}w\; \phi \big( z(u, w)\big) \,  \phi \big( z(v, w)\big), \label{Eq. Xi}
\end{equation}
where we denote
\begin{equation*}
    z(u, w) = (g J) \left( \sqrt{\Delta_0 - |\Delta|} u + \sqrt{|\Delta|} w\right) + g J_0 \, \mu, 
\end{equation*}
$\mathcal{D}\nu = e^{-\nu^2/2}/\sqrt{2\pi} d\nu$ and $\Delta_0 = \Delta(0)$. The functions $\Xi_x$ and $\Xi_y$ thus reduce to: 
\begin{eqnarray} 
\Xi_x(\mathbf{C}; \mathbf{C}_0, \mathbf{M}) &\equiv& \Xi(\Delta_x; (\Delta_x)_0, \mu_x), \\ \Xi_y(\mathbf{C}; \mathbf{C}_0, \mathbf{M}) &\equiv& \Xi(\Delta_y; (\Delta_y)_0, \mu_y), 
\end{eqnarray} 
with $(\Delta_x)_0$ and $(\Delta_y)_0$ fixed.

Because the system of equations in Eq.(\ref{Eq. dynamics C_beta}) is non-conservative, analytical solutions for stationary states are not readily available, as also noted in \cite{kadmon_transition_2015}. Therefore, the stationary values of $\mathbf{M}$ and $\mathbf{C}(\tau)$ must be computed numerically. This is achieved by solving the DMFT equations through an iterative algorithm, as described in \cite{roy_numerical_2019, zou_introduction_2024} and detailed in \ref{App: numerical DMF}.

To analyse transitions between non-quiescent stationary phases, we linearize the dynamics around a reference state and examine the macroscopic response of the mean $\bM$ and autocorrelation $\bC$ \cite{kadmon_transition_2015}. Decaying responses imply linear stability; growing responses signal instability and a transition to a different regime,  yielding a compact $(\beta,\delta)$-dependent stability criterion.

\section{Stability Analysis of Stationary Solutions}
\label{App: ls beta}

In this appendix we derive the phase boundaries between the different stationary solutions by analyzing their linear stability. Stationary states are characterized by $\bM$ and $\bC(\tau)$, satisfing 
equations (\ref{Eq. M_2_beta}) and (\ref{Eq. dynamics C_beta}), respectively. 

In the large time-lag limit, the autocorrelation converges into $
\bC_\infty=\lim_{\tau\to\infty}\bC(\tau)$, defined by 
\begin{equation} \label{Eq. C_inf}
\bC_\infty =
\bXi(\bC_\infty;\bC(0),\bM).
\end{equation}
Thus, the stability of a stationary solution can be studied by linearizing the coupled equations for $(\bM,\bC_\infty)$, defined by Eqs. (\ref{Eq. M_2_beta}) and (\ref{Eq. C_inf}).

Defining the corresponding Jacobian blocks as
\begin{eqnarray}
K_{\alpha\gamma}
&=&
\frac{\partial \Sigma_\alpha}{\partial M_\gamma},
\qquad
U_{\alpha\gamma}
=
\frac{\partial \Sigma_\alpha}{\partial C_{\gamma}},
\nonumber\\
V_{\alpha\gamma}
&=&
\frac{\partial \Xi_\alpha}{\partial M_\gamma},
\qquad
J_{\alpha\gamma}
=
\frac{\partial \Xi_\alpha}{\partial C_{\gamma}},
\end{eqnarray}
with $\alpha,\gamma\in\{x,y\}$, all derivatives being evaluated at the stationary solution $(\bM,\bC_\infty)$. Since $\Sigma_\alpha$ depends on $\bM$ and on the equal-time autocorrelation $\bC(0)$, but not on the asymptotic autocorrelation $\bC_\infty$, one has $\mathbf{U}=\mathbf{0}$. 
Therefore, the Jacobian of the combined system takes a block-triangular form. As a consequence, the stationary solution is linearly stable if
\begin{equation}
\rho(\bK)<1,
\qquad
\rho(\bJ)<1,
\end{equation}
where $\rho(\cdot)$ denotes the spectral radius. Thus, instabilities of the mean activity and of the asymptotic autocorrelation can be analysed independently at the level of the linear stability spectrum.

In principle, in the stability analysis of the full stationary solution $(\bM,\bC_\infty,\bC(0))$, one should also include fluctuations in $\bC(0)$, which are coupled to $\bM$ and $\bC_\infty$. However, it turns out that such fluctuations do not affect the phase boundaries, hence we will not include them here, for the sake of simplicity. We refer the reader interested in the more general stability analysis, including such fluctuations, to the Supplemental Material \cite{martorell_supplemental_nodate}.

\subsection{Destabilization of the Mean Activity}

We first consider instabilities of the stationary mean $\bM$, which control the AC$\to$SC/COA and PA$\to$COA transitions. For
\begin{equation}
z_\alpha=gJ_0\mu_\alpha+gJ\sqrt{\Delta_\alpha(0)}\,u ,
\end{equation}
where $\mu_\alpha=\sum_\beta{\cal N}_{\alpha\beta}M_\beta$, $
\Delta_\alpha(0)=\sum_\beta{\cal M}_{\alpha\beta}C_\beta(0)$, and $u$ is a standard Gaussian variable, the instability of $\bM$ is governed by the Jacobian matrix $\bK$, with elements
\begin{equation}
K_{\alpha\gamma}
=
gJ_0
\left\langle \phi'(z_\alpha)\right\rangle
\big[\mathcal{N}_{\beta, \delta} \big]_{\alpha\gamma}, 
\end{equation}
and $\mathcal{N}_{\beta, \delta}$ defined by Eq. (\ref{Eq. N}). 
The mean activity $\bM$ is stable while the eigenvalue of $\bK$ with largest real part remains smaller than one. The instability threshold is therefore
\begin{eqnarray}
1
&=&
\frac{\Tr\bK}{2}
\,{\rm Re}
\left[
1+
\sqrt{
1-
\left(
\frac{2}{\Tr\bK}
\right)^2
\det\bK
}
\right]
\nonumber\\
&=&
gJ_0 N_1\,
{\rm Re}
\left[
1+\sqrt{1-4(\beta-\delta)N_2}
\right],
\label{Eq:mean_instability}
\end{eqnarray}
where
\begin{eqnarray} \label{Eq. N1 N2}
N_1
&=&
\frac{
\left\langle \phi'(z_x)\right\rangle
-
\delta\left\langle \phi'(z_y)\right\rangle
}{2},
\nonumber\\
N_2
&=&
\frac{
\left\langle \phi'(z_x)\right\rangle
\left\langle \phi'(z_y)\right\rangle
}{
\left(
\left\langle \phi'(z_x)\right\rangle
-
\delta\left\langle \phi'(z_y)\right\rangle
\right)^2
}
\nonumber\\
&=&
\frac{
\left\langle \phi'(z_y)\right\rangle
}{
\left\langle \phi'(z_x)\right\rangle
}
\left(
1-\delta
\frac{
\left\langle \phi'(z_y)\right\rangle
}{
\left\langle \phi'(z_x)\right\rangle
}
\right)^{-2}.
\end{eqnarray}

Equation~\eqref{Eq:mean_instability} gives the mean-activity bifurcation line, corresponding to Eq. (\ref{Eq. chaos_beta 2}). For PA$\to$COA, it is evaluated at the fixed-point solution $\bC(0)=\bq$, with $\bq=\bXi(\bq;\bq,\bM)$ and $\bM$ given by Eq.~\eqref{Eq. M_beta}. For AC$\to$SC/COA, it is evaluated at $\bM=0$, with $\bC(0)$ obtained numerically from the stationary DMFT equations (see \ref{App: numerical DMF}); in this case $\bC(\tau)\to\bC_\infty=0$ as $\tau\to\infty$.

\subsection{Destabilization of the Autocorrelation}

We next consider instabilities of the autocorrelation function $\bC_\infty$, which control the PA$\to$SC transition (where a fixed-point solution becomes unstable to a chaotic stationary state). The corresponding Jacobian matrix $\bJ$ is described by the elements
\begin{equation}
J_{\alpha\gamma}
=
(gJ)^2
\left\langle [\phi'(z_\alpha)]^2\right\rangle
\big[\mathcal{M}_{\beta, \delta} \big]_{\alpha\gamma}.
\end{equation}
and $\mathcal{M}_{\beta, \delta}$ defined by Eq. (\ref{Eq. M}). 
The asymptotic autocorrelation $\bC_\infty$ is stable while $\rho(\bJ)<1$; hence the transition is determined by $\rho(\bJ)=1$. The instability threshold is given by the condition
\begin{eqnarray}
1
&=&
\frac{\Tr \bJ}{2}
\left[
1+\sqrt{
1-
\left(
\frac{2}{\Tr\bJ}
\right)^2
\det\bJ}
\right]
\nonumber\\
&=&
(gJ)^2M_1
\left[
1+\sqrt{1-4(\delta^2-\beta^2)M_2}
\right],
\label{Eq:autocorr_instability}
\end{eqnarray}
with
\begin{eqnarray} \label{Eq. M1 M2}
M_1
&=&
\frac{
\left\langle [\phi'(z_x)]^2\right\rangle
+
\delta^2
\left\langle [\phi'(z_y)]^2\right\rangle
}{2},
\nonumber\\
M_2
&=&
\frac{
\left\langle [\phi'(z_x)]^2\right\rangle
\left\langle [\phi'(z_y)]^2\right\rangle
}{
\left(
\left\langle [\phi'(z_x)]^2\right\rangle
+
\delta^2
\left\langle [\phi'(z_y)]^2\right\rangle
\right)^2
}
\nonumber\\
&=&
\frac{
\left\langle [\phi'(z_y)]^2\right\rangle
}{
\left\langle [\phi'(z_x)]^2\right\rangle
}
\left(
1+\delta^2
\frac{
\left\langle [\phi'(z_y)]^2\right\rangle
}{
\left\langle [\phi'(z_x)]^2\right\rangle
}
\right)^{-2}.
\end{eqnarray}
Equation~\eqref{Eq:autocorr_instability} gives the PA$\to$SC boundary as described by Eq. (\ref{Eq. chaos_beta 1}) and is evaluated on the fixed-point branch, $\bC(0)=\bq$, with $\bq=\bXi(\bq;\bq,\bM)$.

\section{Dynamical Evolution of the Mean Activity}
\label{App. Dynamics Mean}

In this appendix we characterize the linear dynamics of the mean activity close to the AC--COA transition. Averaging Eq.~\eqref{Eq: DMF-beta} gives
\begin{equation}
\partial_t \bM(t)
=
-\bM(t)
+
\bSigma\big(\bM(t);\bC(t,t)\big),
\end{equation}
where $\bSigma$ is defined in Eq.~\eqref{Eq. M_2_beta}. As noted above, in the AC phase $\bM=0$ and the state is stationary. Close to the transition, a linear perturbation $\delta \bM$ of the mean activity evolves following the linear differential equation
\begin{equation}
\partial_t \delta\bM
=
(\bK-\rm{Id})\delta\bM ,
\label{Eq. JMM}
\end{equation}
with $\bK$ defined in Eq.~\eqref{Eq:mean_instability}. Thus, the relaxation or growth of mean perturbations is determined by the eigenmodes of $\bK-\rm{Id}$.

The corresponding eigenvalues are
\begin{equation}
\lambda_M^{\pm}
=
gJ_0 N_1
\left(
1
\pm
\sqrt{1-4(\beta-\delta)N_2}
\right)
-1 ,
\label{Eq:eigenvalues_mean}
\end{equation}
where $N_1$ and $N_2$ are given in Eq.~\eqref{Eq. N1 N2}. The associated eigenvectors can be written, up to normalization, as
\begin{equation}
\mathbf{v}_M^{\pm}
=
\begin{bmatrix}
1\\[4pt]
\dfrac{
gJ_0\left\langle \phi'(z_x)\right\rangle
-1
-\lambda_M^{\pm}
}{
gJ_0\beta\left\langle \phi'(z_x)\right\rangle
}
\end{bmatrix}.
\end{equation}
Hence, a small perturbation evolves as
\begin{equation}
\delta\bM(t)
\sim
\mathbf{v}_M^{\pm}e^{\lambda_M^{\pm}t}.
\end{equation}
When the leading eigenvalues form a complex-conjugate pair, the instability is oscillatory. In that case, the oscillation frequency at onset is given by
\begin{equation}
\omega_0
=
\left|\operatorname{Im}\lambda_M^{\pm}\right|
=
gJ_0N_1
\sqrt{4(\beta-\delta)N_2-1}.
\end{equation}
The AC--COA transition occurs when $\operatorname{Re}\lambda_M^{\pm}=0$.

\section{Numerical resolution of the self-consistency DMFT equations} \label{App: numerical DMF}

To obtain the macroscopic observables \(\bM(t)\) and \(\bC(t,s)\), we solve Eq.~(\ref{Eq: DMF-beta}) numerically by imposing self-consistent statistics~\cite{roy_numerical_2019, zou_introduction_2024}. The idea is to generate effective trajectories \(x(t),y(t)\) drawn from a Gaussian process with prescribed mean \(\bM(t)\) and autocorrelation \(\bC(t,s)\), compute updated statistics \(\bM^{\mathrm{new}}(t)\), \(\bC^{\mathrm{new}}(t,s)\) from those trajectories, and iterate until convergence (equivalently, until the mean and autocorrelation become stable).

\subsection{Description of the algorithm}

Discretize the time interval \([0,T]\) with step \(h\) (so that \(N_T =T/h\in\mathbb{N}\)). Initialize \(x(0)=x_0\), \(y(0)=y_0\) randomly. The iteration proceeds as follows:
\begin{enumerate}
    \item \textbf{Initialize:} Provide initial guesses for the mean activities \(M_x(t),M_y(t)\) and correlation functions \(C_x(t,s),C_y(t,s)\). Choose a relaxation parameter \(\alpha\in(0,1]\) and a convergence tolerance \(\epsilon\).
    \item \textbf{Iterate until convergence or for a maximum of \(N\) outer steps:}
    \begin{enumerate}
        \item Compute the mean input currents \(\mu_x(t),\mu_y(t)\) from Eqs.~(\ref{Eq. mus}) and the noise covariances \(\Delta_x(t,s),\Delta_y(t,s)\) from Eqs.~(\ref{Eq. deltas}).
        \item Generate \(S\) effective trajectories \(\{(x_i^t,y_i^t)\}_{i=1}^S\) for \(t=0,\ldots,N_T\) using an Euler discretization:
        \begin{align}
            x_i^{t+1} &= (1-h)\,x_i^{t} + h\,\tanh\!\big(gJ_0\,\mu_x(t) + gJ\,\eta_{x,i}^{\,t}\big), \nonumber\\
            y_i^{t+1} &= (1-h)\,y_i^{t} + h\,\tanh\!\big(gJ_0\,\mu_y(t) + gJ\,\eta_{y,i}^{\,t}\big), \nonumber
        \end{align}
        where \(\boldsymbol{\eta}_i^{\,t}=(\eta_{x,i}^{\,t},\eta_{y,i}^{\,t})\) is a zero-mean bivariate Gaussian process with temporal covariances prescribed by \(\Delta_x,\Delta_y\) (and cross-covariances if applicable).
        \item Estimate updated statistics by averaging over realizations:
        \[
        M^{\mathrm{new}}_{x}(t)=\frac{1}{S}\sum_{i=1}^S x_i^t,\quad 
        M^{\mathrm{new}}_{y}(t)=\frac{1}{S}\sum_{i=1}^S y_i^t,
        \]
        and similarly \(C^{\mathrm{new}}_{x}(t,s),C^{\mathrm{new}}_{y}(t,s)\).
        \item Apply relaxation:
        \begin{align}
            M_x &\leftarrow (1-\alpha)M_x+\alpha\,M_x^{\mathrm{new}}, \\ 
            M_y &\leftarrow (1-\alpha)M_y+\alpha\,M_y^{\mathrm{new}}, \nonumber\\
            C_x &\leftarrow (1-\alpha)C_x+\alpha\,C_x^{\mathrm{new}}, \\
            C_y &\leftarrow (1-\alpha)C_y+\alpha\,C_y^{\mathrm{new}}. \nonumber
        \end{align}
        \item \textbf{Stopping criterion:} stop if
        \(
        \max\!\{\|M_x-M_x^{\mathrm{new}}\|,\|M_y-M_y^{\mathrm{new}}\|,\|C_x-C_x^{\mathrm{new}}\|,\|C_y-C_y^{\mathrm{new}}\|\}<\epsilon.
        \)
    \end{enumerate}
\end{enumerate}

\textbf{Output:} The final \(M_x,M_y,C_x,C_y\) constitute a self-consistent solution of the DMFT equations (within discretization and sampling error).

\subsection{Comments and limitations}
Although the algorithm is, in principle, formulated without assuming stationarity—macroscopic observables can be defined at generic times \(t\) and \(s\)—its practical use has important limitations. Instabilities arise when macroscopic observables lose self-averaging, as occurs after the bifurcation of the mean activity (\(\bM\neq0\)): by symmetry, both branches (\(\pm\bM\)) are admissible. Non–fixed-point dynamics amplify sample-to-sample fluctuations, driving different realizations to different branches (in contrast, the PA phase remains stable). For this reason, we restricted our simulations to the AC phase, where the scheme is numerically stable.

In the SC phase, symmetry breaking of the mean (\(\bM\neq0\)) destroys self-averaging~\cite{martorell_ergodicity_2025}: different effective trajectories may settle on \(+\bM\) or \(-\bM\), and the iteration becomes unstable. One can regularize this by enforcing a single branch (e.g., via an explicit selection or an infinitesimal bias), but this introduces external artifacts and explicit symmetry breaking.

In the COA phase, macroscopic observables are non-stationary—\(\bM(t)\) and \(\bC(t,s)\) depend explicitly on time—leading to strong sensitivity to initial conditions and, consequently, to a loss of self-averaging.

\section{Kuramoto analysis}
\label{App. Kuramoto}

We quantify the synchronization degree within each population via a Kuramoto order parameter computed from analytic signals. In the following we sketch the main procedure. 

To remove fixed-point components, we center each trace by its temporal mean:
\begin{equation}
\tilde x_k(t)=x_k(t)-\langle x_k(t)\rangle_t,\qquad
\tilde y_k(t)=y_k(t)-\langle y_k(t)\rangle_t 
\end{equation}
where $\langle \cdot \rangle_t$ stands for average over time.

To quantify synchronization, each neuronal activity is mapped to an analytic signal with instantaneous amplitude and phase via the Hilbert transform. For centered activities \(\tilde x_i(t)\) and \(\tilde y_i(t)\),
\begin{equation}
\begin{aligned}
s_x^{\,k}(t) &= \tilde x_k(t) + i\,\mathcal{H}[\tilde x_k](t) \;\equiv\; a_x^{\,k}(t)\,e^{i\varphi_x^{\,k}(t)},\\
s_y^{\,k}(t) &= \tilde y_k(t) + i\,\mathcal{H}[\tilde y_k](t) \;\equiv\; a_y^{\,k}(t)\,e^{i\varphi_y^{\,k}(t)},
\end{aligned}
\end{equation}
where \(\mathcal{H}[\cdot]\) denotes the Hilbert transform, \(a_{x,y}^{\,i}(t)\ge 0\) are instantaneous amplitudes, and \(\varphi_{x,y}^{\,i}(t)\) the instantaneous phases \cite[Ch.~2]{cohen_time_1995}.

We define the (complex-valued) macroscopic observables $Z_x(t)$ and $Z_y(t)$ as the average of the instantaneous phases within each population:
\begin{equation}
    \begin{aligned}
S_x(t) = R_x(t)\,e^{i\Phi_x(t)} = \frac{1}{N}\sum_{k=1}^{N} e^{i\varphi_x^{\,k}(t)} \\
S_y(t) = R_y(t)\,e^{i\Phi_y(t)}=\frac{1}{N}\sum_{k=1}^{N} e^{i\varphi_y^{\,k}(t)}
    \end{aligned}
\end{equation}
where $R_{x,y}(t)$ $\Phi_{x,y}(t)$ denote the instantaneous amplitude and phase of the macroscopic observable. 

The amplitude is bounded, \(R_{x,y}(t)\in[0,1]\), and measures the degree of synchrony within each population: for incoherent dynamics, \(R=0\); while for fully phase-aligned activity, \(R=1\). This observable is known as Kuramoto parameter. 

In order to characterize the feedback between excitatory and inhibitory populations, we measure the instantaneous phase-delay between $S_x(t)$ and $S_y(t)$: 
\begin{equation}
    \Delta \Phi(t) = \Phi_x(t) - \Phi_y(t).
\end{equation}
In simulations, the phase-delay is recorded when a global oscillation is detected; that is, if the power spectrum detects a non-trivial dominant frequency. 

\bibliographystyle{elsarticle-num}
\bibliography{biblioteca.bib}

\clearpage
\includepdf[pages=-]{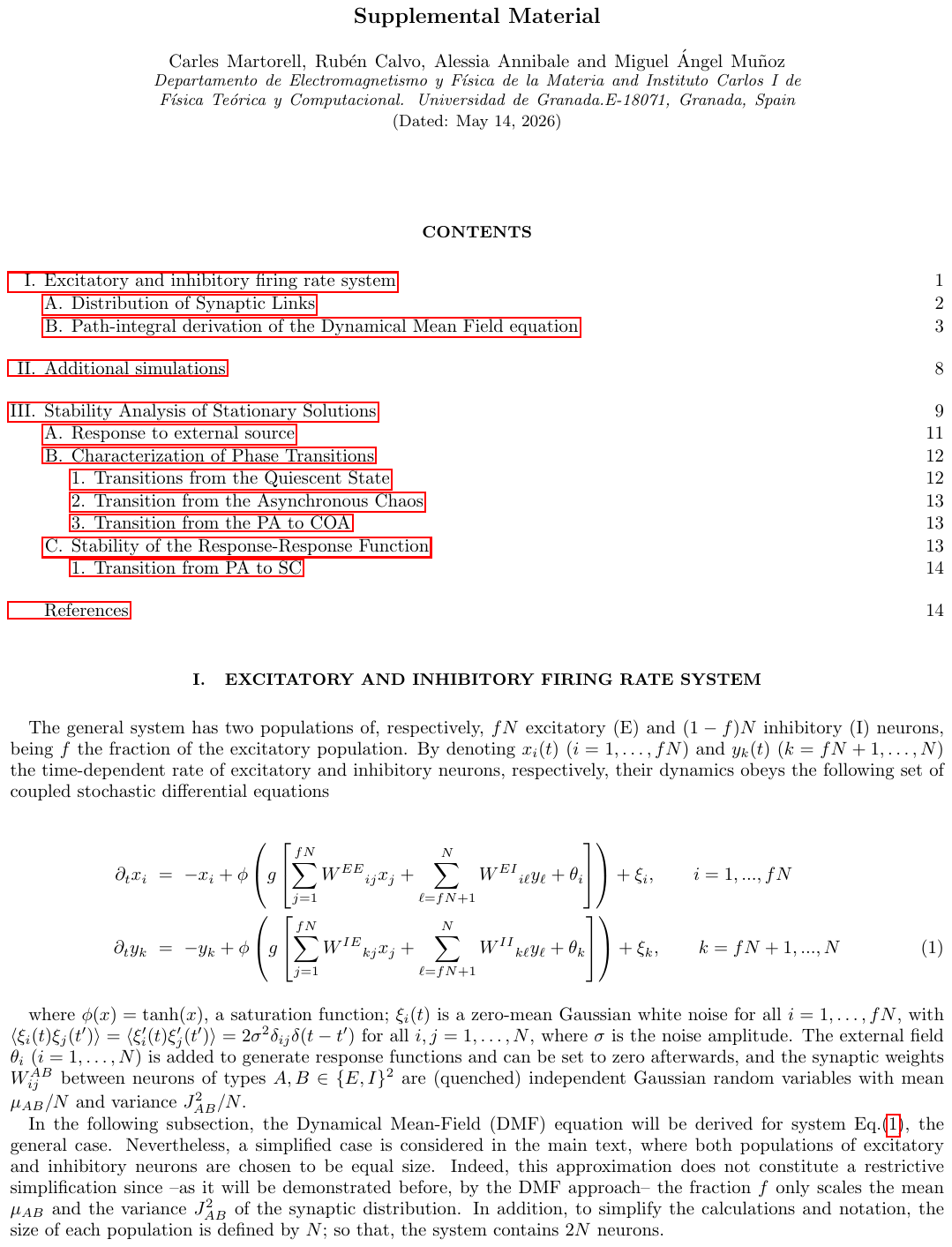}

\end{document}